\documentclass[12pt]{article}
\pdfoutput=1
\usepackage{amssymb,amsmath}
\usepackage{epsfig}
\usepackage{epstopdf}
\usepackage{latexsym}
\usepackage{graphicx}
\usepackage{subfigure}
\usepackage{booktabs}
\usepackage{bbm}
\usepackage[margin=20pt,small]{caption}

\usepackage[toc]{appendix}

\usepackage{color}
\usepackage{datetime}
  
\def\cJ{{\cal J}}

\def\cN{{\cal N}}
\def\cO{{\cal O}}

\def\cL{{\cal L}}

\def\cL{{\cal L}}
\def\cA{{\cal A}}
\def\cB{{\cal B}}

\def\cI{{\cal I}}
\def\cJ{{\cal J}}
\def\cm{{\mathfrak m}}
\def\cphi{\Delta \phi_*}

\definecolor{cardinal}{rgb}{0.6,0,0}
\definecolor{darkgreen}{rgb}{0,0.5,0}
\definecolor{golden}{rgb}{0.92, 0.7, 0}
\definecolor{midnight}{rgb}{0, 0, 0.5}
\definecolor{darkblue}{rgb}{0.2, 0, 0.8}

\begin{document}

\begin{titlepage}

\begin{flushright}
DIAS-STP-13-04 \\
UTTG-15-13 \\
TCC-011-13
\end{flushright}

\bigskip
\bigskip
\bigskip
\centerline{\Large \bf Dynamics of Non-supersymmetric Flavours}
\bigskip
\bigskip
\bigskip
\centerline{{\bf M. Sohaib Alam$^{1}$, Matthias Ihl$^{2}$, Arnab Kundu$^{1}$ and Sandipan Kundu$^{1,3}$}}
\bigskip
\bigskip
\centerline{$^1$ Theory Group, Department of Physics}
\centerline{University of Texas at Austin} 
\centerline{Austin, TX 78712, USA.}
\bigskip
\centerline{$^2$ School of Theoretical Physics}
\centerline{Dublin Institute for Advanced Studies} 
\centerline{10 Burlington Rd, Dublin 4, Ireland.}
\bigskip
\centerline{$^3$ Texas Cosmology Center}
\centerline{University of Texas at Austin} 
\centerline{Austin, TX 78712, USA.}
\bigskip
\centerline{msihl[at]stp.dias.ie, malam, arnab, sandyk[at]physics.utexas.edu}
\bigskip
\bigskip

\begin{abstract}

\noindent{We investigate the effect of the back-reaction by non-supersymmetric probes in the Kuperstein-Sonnenschein model. In the limit where the back-reaction is small, we discuss physical properties of the back-reacted geometry. We further introduce additional probe flavours in this back-reacted geometry and study in detail the phase structure of this sector when a constant electromagnetic field or a chemical potential are present. We find that the Landau pole, which serves as the UV cut-off of the background geometry, also serves as an important scale in the corresponding thermodynamics of the additional flavour sector. We note that since these additional probe flavours are indistinguishable from the back-reacting flavours, the results we obtain point to a much richer phase structure of the system.}

\end{abstract}

\newpage

\tableofcontents

\end{titlepage}

\newpage

\section{Introduction}

Use of the AdS/CFT correspondence\cite{Maldacena:1997re, Witten:1998qj, Gubser:1998bc} has provided us with a remarkably powerful set of techniques to address strong coupling dynamics of certain gauge theories. Over the years it has become possible to capture some qualitative but key features of Quantum Chromodynamics (QCD), which so far has eluded the standard lore of conventional field theory methods.

In this article we will discuss one such aspect: the physics of the chiral symmetry breaking within the flavour sector of a large $N_c$ gauge theory, with the hope that we learn at least qualitatively useful lessons about QCD and therefore the strongly coupled matter created at the Relativistic Heavy Ion Collider (RHIC) and the Large Hadron Collider (LHC). We will rely on a model proposed in \cite{Kuperstein:2008cq}\footnote{See also \cite{Dymarsky:2009cm}.}, which we will henceforth refer to as the Kuperstein-Sonnenschein model.

The standard way of discussing the flavour physics is to introduce a set of probe branes in a particular gravity background which is sourced by some stack of D-branes. This method was pioneered in \cite{Karch:2002sh}. In the Kuperstein-Sonnenschein model, a pair of probe D7 and anti-D7 brane is placed in the AdS$_5 \times T^{1,1}$ background which is popularly known as the Klebanov-Witten background\cite{Klebanov:1998hh}. The probe branes wrap a three cycle inside the internal manifold $T^{1,1} \cong S^2 \times S^3$ and extend along the rest of the conifold ${\mathbb R}^+ \times S^2$. The zero temperature physics of the probes realizes a spontaneous breaking of the chiral symmetry by having the branes join smoothly in the IR and thus leading to: $U(N_f)_L \times U(N_f)_R \to U(N_f)_{\rm diag}$. In \cite{Alam:2012fw}, we have explored the dynamics of the flavours in this model in the presence of a temperature and an external electro-magnetic field and found an interesting and rich phase structure.

The physics explored in the probe limit teaches us interesting lessons about the dynamics of flavours in such models, however they are also limited by the probe approximation. It is an interesting question in its own right to consider going beyond this approximation. Such an exercise is also physically relevant for QCD, where the number of colours and the number of flavours are of the same order. To consider back-reaction by the probes on the background geometry, one has to solve for the equations of motion obtained from an action consisting of supergravity and Dirac-Born-Infeld pieces. Typically, to facilitate the technical challenges, such an undertaking is carried out within the so called ``smearing" approximation, where the probes are smeared along their transverse directions such that the full symmetry of the original background is recovered.

Work along this direction has been carried out in \cite{Bigazzi:2005md}-\cite{Caceres:2007mu} in related models and summarized in the review \cite{Nunez:2010sf}. However, most of these efforts rely on supersymmetry and become technically simpler than solving the equations of motion. The Kuperstein-Sonnenschein model is non-supersymmetric to begin with and thus one needs to consider a system of coupled second order non-linear differential equations to make any progress. We explored this system of equations in \cite{Ihl:2012bm} and found an analytic solution at the leading order in the back-reaction, measured in powers of $N_f/N_c$.

When the flavour back-reaction is taken into account, the underlying conformal theory is deformed by higher dimensional irrelevant operators of dimension $6$ and dimension $8$ respectively. Furthermore, the gauge coupling runs and the theory acquires a Landau pole in the UV: the resulting theory is not UV-complete. Interestingly, these generic properties are very similar to the back-reacted backgrounds which do preserve some supersymmetry. In \cite{Ihl:2012bm} we have also demonstrated that an additional probe sector in this back-reacted background now undergoes a chiral phase transition, which is otherwise absent when the back-reaction vanishes. This is simply because the Landau pole now gives a scale and there is a clear notion of small and large temperature regimes.

Let us offer a few more comments. It is not possible to observe a finite temperature phase transition in a theory whose underlying description is conformal. However, if one introduces another scale in the system (other than the temperature), as we have explored in \cite{Alam:2012fw}, this possibility opens up. Typically, any such scale is an ``infrared property" of the system since we are probing the system at low energies, {\it e.g.} at room temperatures. This is often summarized by saying that thermodynamics is an infrared property of a system, which means it does not care about the energy-scales that lie much higher than what is explored in an experimental set-up.

On the other hand, we have an example, demonstrated in \cite{Ihl:2012bm}, where such a phase transition happens because the Landau pole provides a scale other than the temperature. The theory is valid well below the Landau pole; and we have shown in \cite{Ihl:2012bm} that within the regime of its validity there is now a phase transition. Thus the thermodynamics of the system is not quite just an ``infrared property" of the system anymore. If we had an UV-complete description of the back-reacted system, the Landau pole would disappear yielding a more fundamental microscopic description which is not conformal. Thus it is not a surprising fact that a finite temperature phase transition takes place in such a fundamentally non-conformal system.

Before proceeding further, let us remark on a possible limitation\footnote{We thank Jacques Distler and Vadim Kaplunovsky for raising this point.} of our approach as far as exploring the physics in the flavour sector is concerned. We will introduce an additional probe sector and discuss the phase structure of this additional probe sector in the back-reacted background. Note, however, that our back-reacted background is obtained by considering back-reaction by the parallel shaped embeddings only. Thus, if we observe a phase transition from the parallel-shaped to the U-shaped embeddings in this additional probe sector, perhaps this is an indication that we need to actually consider back-reaction by both parallel-shaped and the U-shaped profiles separately and then compare the free energies of these two backgrounds. This is a very interesting yet technically more involved problem subject to ongoing research. For present purposes, we will adhere to a simpler analysis and pretend that the additional probe sector and the back-reacting probe sector are distinguishable, which should be viewed as a first attempt towards exploring the actual issue.

In the present article we will explore more aspects of the flavour sector by introducing additional probe branes in the back-reacted background and exciting a constant electro-magnetic field on this additional probe system. Our focus is to study the flavour dependence on the phase structure obtained in \cite{Alam:2012fw}. This effort, the reader should note, is merely a first attempt to understand how the QCD phase diagram might depend on the number of flavour degrees of freedom. Using lattice simulations, current understanding of flavour dependence of the QCD phase diagram at vanishing chemical potential is usually summarized in the so called ``Columbia plot" (see for example fig 1 in \cite{Bonati:2012pe}).\footnote{We thank Massimo D'Elia for very useful correspondence and the reference.} However, there is no general understanding about the phase structure in the presence of external fields such as the ones considered in this article. Thus our hope is to learn about robust qualitative features which may be relevant to QCD; although we have to remember that since we account for the back-reaction only up to the leading order in $N_f/N_c$, the flavour dependence that we will find within this framework is, by definition, weak. Nonetheless, the fact that this back-reaction breaks the conformal invariance of the background will result in some drastic changes as compared to the case when the back-reaction vanishes. 

This article is divided in the following sections: Section 2 reviews the most relevant results on the back-reacted background obtained in \cite{Ihl:2012bm}. In section 3, we discuss some interesting physical aspects of the perturbative back-reacted solution, while in section 4, electro-magnetic fields are introduced in the probe sector and their effects are investigated thoroughly, including their impact on holographic renormalization of UV divergences of the on-shell DBI action. Moreover, section 4 contains an exhibition of the most pertinent effects associated with the introduction of a chemical potential, studied both in the grand-canonical and canonical ensembles. Finally, section 5 offers concluding remarks and an outlook on future research.

\section{Review of previous results}

\subsection{The back-reacted background}

Let us begin with a brief review of the earlier results based on which we will continue to explore similar physical effects in our current work. Before taking any back-reaction into account, the model we consider is described in \cite{Kuperstein:2008cq}. The authors introduced probe D7/anti-D7 branes in the AdS$_5\times T^{1,1}$ background, which is obtained as the near-horizon limit of a stack of D3-branes placed at the tip of the conifold. The D7/anti-D7 branes wrap a $3$-cycle in the internal manifold $T^{1,1}\cong S^2\times S^3$ and are extended along the rest of the conifold.

Before introducing the flavours, the dual field theory is given by an $\cN=1$ superconformal quiver gauge theory in $(3+1)$-dimensions with a gauge group $SU(N_c)\times SU(N_c)$ and a global $SU(2)\times SU(2) \times U(1)_R$ symmetry group. The degrees of freedom are contained in two bi-fundamental chiral superfields which transform in the $\left(N_c, \bar{N_c}\right)$ and $\left(\bar{N_c}, N_c\right)$ representations of the gauge group.

Introducing the probe branes corresponds, in the dual field theory, to introducing flavour degrees of freedom in an analogue of the so called ``quenched approximation". This amounts to introducing a global $U(N_f)_L \times U(N_f)_R$ flavour symmetry group, where $N_f$ denotes the number of flavours. The zero temperature physics of this system captures a geometric realization of the spontaneous breaking of chiral symmetry: the brane--anti-brane pair joins in the IR breaking the aforementioned flavour symmetry group down to a diagonal $U(N_f)$. On the other hand, the finite temperature physics of this system is rather trivial: since the background is conformal, there is no scale in the system and hence no phase transition can happen. Chiral symmetry is always restored in this case \cite{Alam:2012fw}. Nonetheless, the system exhibits interesting phase structure and some interesting phenomenon, such as the effect of {\it magnetic catalysis} in chiral symmetry breaking, when more control parameters are introduced \cite{Alam:2012fw}. In this article we will analyze the effect of the back-reaction by flavours on the physics observed and analyzed in \cite{Alam:2012fw}.

Towards that end, we need to find the back-reacted background. Such a background can be found by sourcing the supergravity equations of motion by the Dirac-Born-Infeld (DBI) contribution coming from the probe flavour degrees of freedom. It turns out that, employing the ``smearing technique", we can find an analytical solution of these back-reacted equations of motion at the leading order in the $N_f/N_c$ correction \cite{Ihl:2012bm}. In Einstein frame, the most general form of the back-reacted background is given by
\begin{eqnarray} \label{backmet}
ds^2 & = & h(r)^{-1/2} \left(- b(r) dt^2 + d \vec{x}^2 \right) + h(r)^{1/2} \left[ \frac{dr^2}{b(r)} + \frac{e^{2g(r)}}{6} \sum_{i=1,2} \left( d\theta_i^2 + \sin^2 \theta_i d\phi_i^2 \right) \right. \nonumber\\
& + & \left.  \frac{e^{2f(r)}}{9} \left(d\psi + \sum_{i=1,2} \cos\theta_i d\phi_i \right)^2 \right] \ , \\
F_5 & = & k(r) h(r)^{3/4} \left( e^t \wedge e^{x^1} \wedge e^{x^2} \wedge e^{x^3} \wedge e^r + e^{\psi} \wedge e^{\theta_1} \wedge e^{\phi_1} \wedge e^{\theta_2} \wedge e^{\phi_2} \right) \ ,
\end{eqnarray}
where the vielbeins are given by
\begin{eqnarray}
&& e^t = h^{-1/4} b^{1/2} dt \ , \quad e^{x^i} = h^{-1/4} dx^i \ , \quad e^r = h^{1/4} b^{-1/2} dr \ , \\
&& e^{\psi} = \frac{1}{3} h^{1/4} e^f \left( d\psi + \cos\theta_1 d\phi_1 + \cos\theta_2 d\phi_2 \right) \ , \\
&& e^{\theta_{1,2}} = \frac{1}{\sqrt{6}} h^{1/4} e^g d\theta_{1,2} \ , \quad e^{\phi_{1,2}} = \frac{1}{\sqrt{6}} h^{1/4} e^g \sin\theta_{1,2} d\phi_{1,2} \ .
\end{eqnarray}
Here $k(r)$ is a function that we can determine from the relation
\begin{eqnarray} \label{radius}
k(r) h(r)^2 e^{4g(r) + f(r)} = 27 \pi g_s^*N_c l_s^4 = 4 L^4 \ ,
\end{eqnarray}
where $g_s^*$ is the string coupling defined at $r=r_*$, which is a UV cut-off that we need to introduce since we have a running dilaton; $l_s$ is the string length and $L$ is the AdS-radius. The various metric functions are given by
\begin{eqnarray}
&& b(r) = 1- \frac{r_H^4}{r^4} \ , \label{solb}\\
&& \Phi(r) = \frac{\epsilon}{4} \log\left( \frac{r}{r_*}\right) \ , \\
&& h(r) = \frac{L^4}{r^4} \left( 1 + \frac{\epsilon}{8} \right) + \epsilon \alpha \left(2 - \frac{r_H^4}{r^4} \right) \ , \\
&& e^{f(r)} = r \left[ 1 + \epsilon \left( -\frac{1}{24} + 4 \cm^{-2} \left(\frac{r_H^4}{r^2}\right) K \left( 1 - \frac{r_H^4}{r^4}\right) - 8 \cm^{-2} r^2 E \left(1 - \frac{r_H^4}{r^4}\right) \right)\right] \ , \\
&& e^{g(r)} = r \left[ 1 + \epsilon \left( -\frac{1}{48} -  \cm^{-2} \left(\frac{r_H^4}{r^2}\right) K \left( 1 - \frac{r_H^4}{r^4}\right) + 2 \cm^{-2} r^2 E \left(1 - \frac{r_H^4}{r^4}\right) \right)\right] \ , \label{solg}
\end{eqnarray}
with
\begin{eqnarray}
\epsilon = \frac{3}{2\pi^2} \left(\frac{\lambda N_f}{N_c}\right) \ .
\end{eqnarray}
Here $r_H$ is the location of the event-horizon, $\alpha$ and $\cm$ are two constants which correspond to --- in the dual field theory --- the couplings of a dimension $6$ and a dimension $8$ operator, respectively.\footnote{Note that the constant $\alpha$ is denoted by $c_3$ in \cite{Ihl:2012bm}. The general solution in \cite{Ihl:2012bm} contains another constant which is denoted by $c_5$; however, since the coupling corresponding to the dimension $8$ operator is a linear combination of these two constants, we have set $c_5 = 0$ without the loss of any generality.} Note that the relevant physical coupling should be defined at the temperature scale. However, for our background this relation takes the form
\begin{eqnarray}
\lambda_{\rm th} = \lambda \left[ 1 + \frac{\epsilon}{4} \log\left(\frac{r_H}{r_*}\right)\right]  + \cO(\epsilon^2) \ , \label{lambdath}
\end{eqnarray}
which yields $\epsilon_{\rm th} = \epsilon + \cO(\epsilon^2)$. Since the background in (\ref{solb})-(\ref{solg}) is valid up to $\cO(\epsilon)$, we have $\epsilon_{\rm th} = \epsilon$.

The temperature of the background can be identified with the inverse period of the Euclidean time direction. This yields \cite{Ihl:2012bm},
\begin{eqnarray} \label{temp}
T = \frac{r_H}{\pi L^2} \left( 1 - \frac{\epsilon}{16} - \frac{\epsilon \alpha}{2} \frac{r_H^4}{L^4}\right) \ .
\end{eqnarray}

Before proceeding further, let us comment on the regime of validity of the solution given in (\ref{solb})-(\ref{solg}). To avoid the Landau pole coming from the diverging dilaton field we need to impose $r_* \ll \infty$, where $r_*$ is the UV cut-off.  Now the perturbative solution for various other functions will hold provided
\begin{eqnarray}
&& \epsilon \left | \log\left(\frac{r_H}{r_*}\right)\right | \ll \cO(1) \ , \quad \epsilon \left | \mathfrak{m}^{-2} \left(\frac{r_H^4}{r_*^2 }\right)\right | \ll \cO(1) \ , \quad \epsilon \left | \mathfrak{m}^{-2} r_*^2\right | \ll \cO(1) \ , \label{cond1} \\
&& {\rm and} \quad 2 \epsilon \left |\alpha \left(\frac{r_*^4}{L^4}\right)\right | \ll \cO(1) \ . \label{cond2}
\end{eqnarray}
%

\subsection{Introducing an additional probe sector}

In \cite{Ihl:2012bm}, we have explored the phase diagram of an additional probe sector in the background given in (\ref{backmet}) and (\ref{solb})-(\ref{solg}) imposing $\alpha =0$, {\it i.e.} setting the source for the dimension $8$ operator to zero. The inclusion of the back-reaction breaks the conformal invariance of the background and we found that the additional probe sector now undergoes a chiral phase transition. This phase transition is, in a very precise sense, caused by the existence of the Landau pole: this pole is located at $r \to \infty$ at the leading order in $\epsilon$, which means we need to use a momentum UV cut-off. In what follows, we will discuss the phase structure in the additional probe sector including more control parameters such as a constant electro-magnetic field.

Before going further, let us revisit the phase structure in more details. Following \cite{Ihl:2012bm}, we introduce $N_f'$ additional probe D7 and anti-D7 branes such that $N_f' \ll N_f$. These $N_f'$ probes are aligned in a way similar to the back-reacting flavour branes. The dynamics of the $N_f'$ flavours are given by
\begin{eqnarray}
S_{\rm DBI} = \cN_T \int dr e^{f(r) + 2g(r) + \Phi(r)} \sqrt{1 + \frac{1}{6} b(r) e^{2g(r)} \phi'^2 } \ , \quad \cN_T =  N_f' \tau_7 V_{\mathbb R^3} \frac{8\pi^2}{9T} \ , \label{actd7}
\end{eqnarray}
where $\tau_7$ is the tension of the probe, $V_{\mathbb R^3}$ denotes the volume along the three spatial directions and $T$ is the background temperature. We have denoted $\phi \equiv \phi_1$ and also used the fact that it is consistent to set $\theta_1 = \pi/2$.

The equation of motion admits two classes of solutions: the parallel-shaped solutions denoted by
\begin{eqnarray}
\phi(r) = {\rm const} \ ,
\end{eqnarray}
and the U-shaped solutions given by
\begin{eqnarray}
\phi'(r) = \frac{6 c}{\sqrt{b(r)^2 e^{2f(r) + 8g(r) + 2 \Phi(r)} - 6 c^2 b(r) e^{2g(r)}}}  \ , \quad c^2 = \left. \frac{1}{6} b(r) e^{2f(r) + 6g(r) + 2\Phi(r)} \right |_{r=r_0} \ ,
\end{eqnarray}
where $r_0$ is the point where the brane--anti-brane pair joins smoothly. The parallel-shaped solutions correspond to the chiral symmetry restored phase and the U-shaped solutions correspond to the chiral symmetry broken phase.

\begin{figure}[h!]
\centering
\includegraphics[scale=0.65]{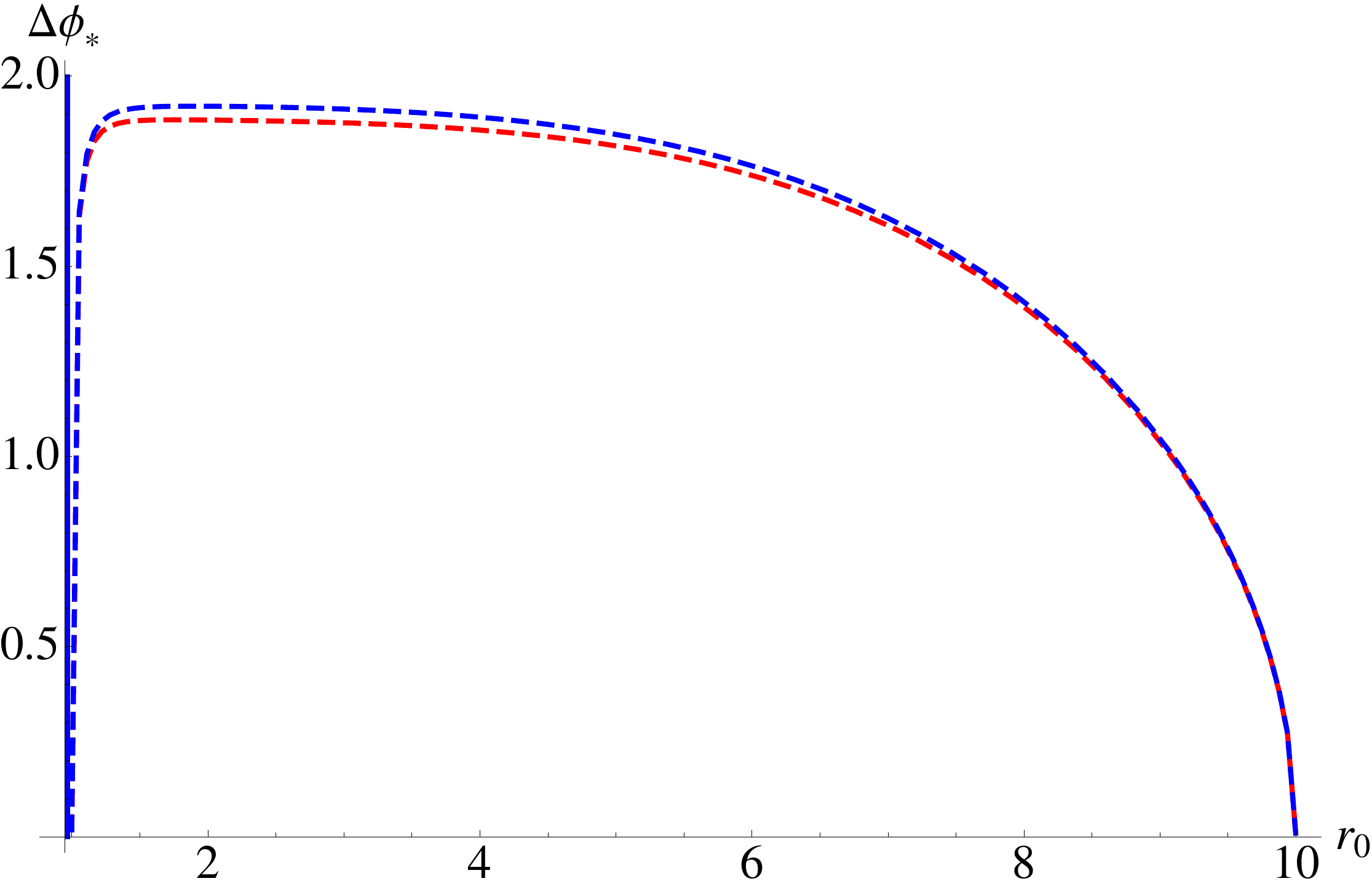}
\caption{\small The blue and the red dashed curves correspond to $\epsilon = 0.01$ and $\epsilon=0.5$ respectively corresponding the U-shaped embeddings. We have set $\mathfrak{m}^{-2}=0$ and $r_H=1$ and $r_*=10$. The solid vertical blue line correspond to the parallel embeddings.}
\label{phir0}
\end{figure}

Before discussing the phase transition, let us comment on the dependence of the coupling $\cphi$ on the expansion parameter $\epsilon$. For $\mathfrak{m}^{-1} = 0 = \alpha$, the asymptotic angular separation is given by
\begin{eqnarray}
\cphi (\epsilon) = \cphi(0) + \epsilon \, \cI(r_0, r_H) \ , \quad {\rm where} \quad \cI(r_0, r_H)< 0 \ . 
\end{eqnarray}
Thus at the leading order in $\epsilon$ the angular separation decreases linearly with $\epsilon$ with a slope which is determined by $r_0$ and $r_H$. The dependence of $\cphi$ with $\epsilon$ is schematically shown in fig.~\ref{phir0}. It is clear that, for the U-shaped profiles, increasing the effect of the back-reaction reduces the maximum value attained by the angular separation, denoted by $\cphi^{\rm max}$.
In fact there is more physics in fig.~\ref{phir0}: it displays the various available phases of the system for a given value of the coupling $\cphi$. For any $\cphi > \cphi^{\rm max}$, we only have the chiral symmetry restored phase available. On the other hand, for any $\cphi < \cphi^{\rm max}$, we have three available points in the phase space: one on the vertical solid line and two on the dashed curves. Comparing this situation with the more familiar $\{P-V\}$-diagram of the Van der Waals gas, we can conclude that there exists a first order phase transition from a point on the solid vertical line ({\it i.e.} the chiral symmetry restored phase) to a point on the dashed line ({\it i.e.} the chiral symmetry broken phase) for some critical value of $\cphi$.

\begin{figure}[h!]
\centering
\includegraphics[scale=0.78]{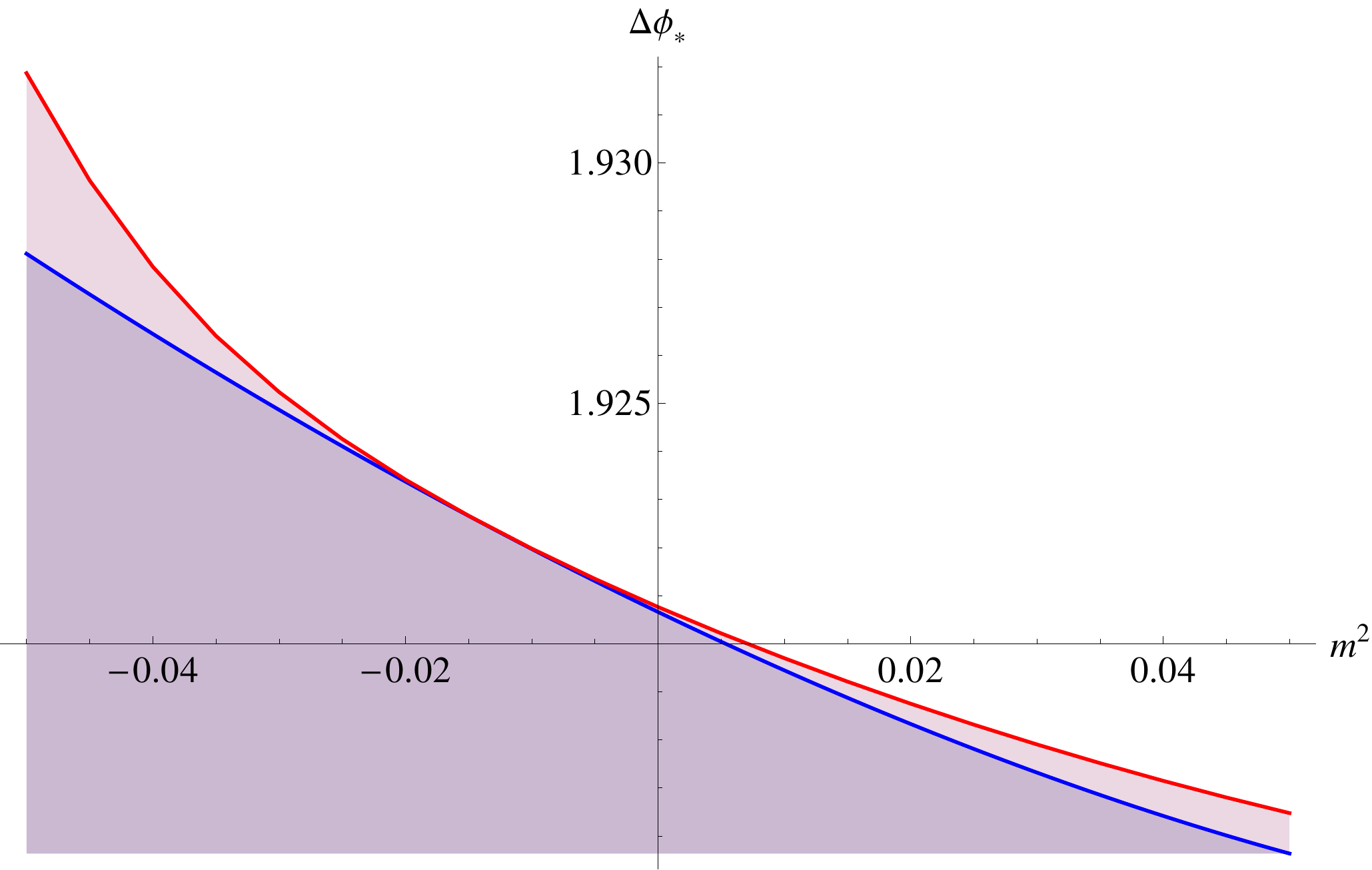}
\caption{\small The phase diagram in the $\cphi$ vs $m^2$ plane. The blue line separates the $\chi$SB $\equiv$ chiral symmetry broken phase (below the line) from the $\chi$SR $\equiv$ chiral symmetry restored phase (above the line). Between the blue and the red line, we also have a metastable $\chi$SB phase. Above the red line we have only the $\chi$SR phase. }
\label{figphase}
\end{figure}

The energetics of the two classes of embeddings will now decide the phase of the system. One has to look at the Euclidean on-shell action --- which is identified with the thermodynamic free energy of the system --- for these two types of embeddings and compute their difference
\begin{eqnarray}
\Delta S = S_U - S_{||} \ .
\end{eqnarray}
Now depending whether $\Delta S>0$ or $\Delta S<0$, we will get a chiral symmetry restored or a chiral symmetry broken phase. This results in a non-trivial phase structure analyzed in \cite{Ihl:2012bm}, which we have shown in fig.~\ref{figphase}. The phase diagram is presented in the $\cphi$ vs $m^2$ plane, where $m^2$ is defined as
\begin{eqnarray}
m^2 = \pi L^4 \left(\frac{T}{\mathfrak m}\right)^2 \left(1 + \frac{\epsilon}{8}\right) \ , \quad {\rm with} \quad \alpha = 0\ .
\end{eqnarray}
We want to emphasize two main features here: First, in the absence of the back-reaction, no such phase structure exists at finite temperature. The underlying theory is conformal and only the chiral symmetry restored phase exists. Taking back-reaction into account breaks the conformal invariance by introducing a UV Landau pole and also deforming the CFT by higher dimensional irrelevant operators. In the leading order back-reacted solution described here, the UV Landau pole is located at $r \to \infty$ and the couplings corresponding to the irrelevant operators of dimension $6$ and $8$ respectively are denoted by $\alpha$ and $\mathfrak{m}$.

From the phase diagram shown in fig.~\ref{figphase} it is clear that $m^2 = 0$ is not a special point as far as the existence of the phase transition is considered. Thus we can conclude that the non-trivial phase structure and the associated thermodynamics that we obtain in this back-reacted model is caused by the existence of the Landau pole. We will now move on to discuss the effect of the back-reaction on some bulk properties of the background as well as the phase structure of an additional probe sector in the spirit of \cite{Alam:2012fw}.

\section{The back-reacted background: some physical aspects}

Let us briefly comment on a few physical properties that we can extract from the solution in (\ref{solb})-(\ref{solg}). To begin with, let us focus on the physics of the energy loss of the plasma. In the presence of the black hole, {\it i.e.}, introducing a non-zero temperature in the dual field theory, dissipation will occur due to the presence of the black hole in the bulk. We can now investigate how the back-reaction of the flavours affect the physics of dissipation at least when the back-reaction is taken into account perturbatively. Within such a ``stringy" framework, there are two\footnote{See also \cite{CasalderreySolana:2006rq, Gubser:2008as}.} canonical ways to explore this: First, in a perturbative description, the mechanism of energy loss of a parton moving in a plasma is usually characterized by the so called jet quenching parameter \cite{Baier:1996sk}, usually denoted by $\hat{q}$. In \cite{Liu:2006ug}, it was proposed that a non-perturbative description of the jet quenching parameter is given by a light-like Wilson loop, which one can use to perform computations at strong coupling. Second, within the ``stringy" framework, the energy loss of a moving quark (parton) can be easily modeled by considering a fundamental string moving with a constant velocity. The force required to drag the string at a constant velocity is essentially the energy that is being lost in the plasma. This was first proposed and explored in \cite{Herzog:2006gh, Gubser:2006bz}. Here we will explore both these cases.

\subsection{Jet quenching parameter}

Let us begin with the jet quenching parameter. Using the general formula in \cite{Armesto:2006zv}, the jet quenching parameter is given by
\begin{eqnarray}
\hat{q}^{-1}  = \pi \alpha' \int_{r_H}^{r_*} e^{-\Phi/2} \frac{\sqrt{G_{rr}} dr}{G_{xx} \sqrt{G_{xx} + G_{tt}}} \ ,
\end{eqnarray}
where $G$ denotes the spacetime metric in the Einstein frame and the dilaton enters because of the conversion factor between the string frame and the Einstein frame. This gives
\begin{eqnarray}
 \hat{q}^{-1} & = & I_0 + \epsilon I_1 + \epsilon \alpha I_2  \ , \quad {\rm where} \nonumber\\
I_0 & = & \alpha' \frac{L^4}{r_H^3} \int_{1}^{r_*/r_H}  \frac{d\rho}{\sqrt{\rho^4 - 1}} \ , \\
I_1 & = & \alpha' \frac{L^4}{8 r_H^3} \int_{1}^{r_*/r_H}  \frac{d\rho}{\sqrt{\rho^4 - 1}} \left[1 - \log\left(r_H/r_* \right) - \log\rho \right] \ , \\
I_2 & = & \alpha' r_H \int_{1}^{r_*/r_H}  \frac{d\rho}{\sqrt{\rho^4 - 1}} \left[ 2 \rho^4 - 1 \right] \ .
\end{eqnarray}
Thus the parameter $\hat{q}$ receives a correction of the form
\begin{eqnarray}
\hat{q} = I_0^{-1} - \epsilon \frac{I_1}{I_0^2} - \epsilon \alpha \frac{I_2}{I_0^2} \ .
\end{eqnarray}
In terms of $\lambda_{\rm th}$, which was defined in (\ref{lambdath}), $\hat{q}^{-1}$ can be re-written as
\begin{eqnarray}
\hat{q}^{-1} & = & \bar{I}_0 + \epsilon \bar{I}_1 + \epsilon \alpha \bar{I}_2 \ , \quad {\rm where} \nonumber\\
\bar{I}_0 & = & \frac{4}{3\sqrt{3} T^3 \pi^3 \sqrt{\lambda_{\rm th}}}  \int_{1}^{r_*/r_H}  \frac{d\rho}{\sqrt{\rho^4 - 1}} \ , \\
\bar{I}_1 & = & \frac{1}{6\sqrt{3} T^3 \pi^3 \sqrt{\lambda_{\rm th}}} \int_{1}^{r_*/r_H}  \frac{d\rho}{\sqrt{\rho^4 - 1}} \left[-\frac{1}{2}  - \log\rho \right] \ , \\
\bar{I}_2 & = & \frac{4\pi L^4 T}{3\sqrt{3} \sqrt{\lambda_{\rm th}}}  \int_{1}^{r_*/r_H}  \frac{d\rho}{\sqrt{\rho^4 - 1}} \left[ 2 \rho^4 - \frac{5}{2} \right] \ .
\end{eqnarray}
Let us now offer a few comments: First, notice that the integrals $\bar{I}_0, \bar{I}_2 >0$, $\bar{I}_1<0$. The sign of the parameter $\alpha$, which is hitherto unconstrained, determines whether the flavour contribution to the jet quenching parameter is positive or negative. The $\alpha$-independent contribution is always positive. Note that, in \cite{Bigazzi:2009bk} the parameter $\alpha$ is set to zero and thus the flavour contribution turns out to be always positive. The inclusion of the back-reaction breaks the underlying conformal symmetry by introducing a dimension $8$ operator, which {\it a priori} can have both positive or negative contribution. The dimension $6$ operator, however, does not play any role here.

\subsection{Drag force computation}

We will keep our discussion brief and follow \cite{Herzog:2006se} closely as far as notations are concerned. To discuss the drag force computation, we need to consider the following: let us start from the back-reacted background and consider a string that is hanging from the boundary, which in this case is located at $r=r_*$. The end point of the string describes a quark moving in the plasma as the string propagates in the bulk space-time. The ansatz for such a moving string takes the form
\begin{eqnarray}
x \left(\sigma, \tau\right) = r \left(\sigma \right) + v \tau \ ,
\end{eqnarray}
where $\{\tau, \sigma\}$ represents the string worldsheet and $v$ denotes the velocity of the quark.

Due to the constant velocity, the string worldsheet develops a horizon that can be obtained from the following formula:
\begin{eqnarray}
\left. G_{tt} + G_{xx} v^2 \right|_{r=r_c} = 0  \ , \quad \implies \quad r_c = r_H \left(1 - v^2 \right)^{-1/4} \ ,
\end{eqnarray}
where $r_c$ denotes the location of the horizon and $G$ once again denotes the background spacetime. The momentum transfer is given by \cite{Herzog:2006se}
\begin{eqnarray}
\frac{dp}{dt} =  - \frac{1}{2\pi} C \ , \quad {\rm with} \quad \left. e^{\Phi} G_{tt} G_{xx} + \alpha'^2 C^2\right|_{r=r_c} = 0 \ ,
\end{eqnarray}
where the dilaton $\Phi$ enters the above formula since $G$ is presented in the Einstein frame. Thus we finally get
\begin{eqnarray}
C & = & \frac{L^2 \pi^2 T^2 v}{\alpha' \sqrt{1-v^2}} \left[ 1 + \frac{\epsilon}{16} + \frac{\epsilon}{8} \log \left(\frac{L^2 \pi T }{r_* \left( 1 - v^2 \right)^{1/4}}\right) + \frac{\epsilon}{2} L^4 \pi^4 T^4 \alpha \frac{1- 3v^2}{1-v^2}\right] \nonumber\\
& = & \frac{3\sqrt{3}}{4} \frac{\pi^2 T^2 v}{\sqrt{1-v^2}} \sqrt{\lambda_{\rm th}} \left[ 1 + \frac{\epsilon}{16} - \frac{\epsilon}{8} \log\left(1 - v^2 \right)^{1/4} + \frac{\epsilon\alpha}{2} L^4 \pi^4 T^4 \frac{1- 3 v^2}{1 - v^2}\right] \ ,
\end{eqnarray}
where we have used the 't Hooft coupling defined at the temperature scale from (\ref{lambdath}). From the second line above we observe that for $\alpha (1- 3 v^2) \ge 0$, the energy loss is enhanced by the presence of the back-reaction. Note that $v^2=1/3$ corresponds to the speed of sound for the medium and therefore it is curious to observe that the energy loss changes signature across this speed.

\subsection{Quark--anti-quark potential}

We can use the background found in (\ref{solb})-(\ref{solg}) to explore some of the non-perturbative aspects of the dual theory. The dual theory in this case is a non-supersymmetric theory which consists of the Klebanov-Witten theory coupled with a chiral flavour sector. We can investigate how the interaction between the flavours is affected by the presence of the adjoint as well as the fundamental degrees of freedom in this non-supersymmetric theory. To explore this, we can consider a massive quark anti-quark pair such that the mass of the pair is very small compared to the Landau pole. The corresponding bound state 	is given by a string worldsheet ending on a probe flavour brane that extends from $r_0$ to $r_*$, where $r_0$ is some infrared scale satisfying $r_0 \ge r_H$.

To parametrize the worldsheet of the string we can choose: $\tau = t, \sigma = x^1$ and $r= r(x^1)$, where $x^1$ is one of the spatial directions ranging from $-\ell/2$ to $+\ell/2$; and $\{\tau, \sigma\}$ represents the string worldsheet parameters. The quark--anti-quark distance, denoted by $\ell$, and the renormalized potential, denoted by $V$, are then given by\cite{Rey:1998ik, Brandhuber:1998er}
\begin{eqnarray}
&& \ell \left(r_0 \right) = 2 \int_{r_0}^{r_*} \frac{G P_0} {P \sqrt{P^2 - P_0^2}} dr \ , \label{lenqq} \\
&& V \left(r_0 \right) = \frac{1}{\pi \alpha'}  \left[ \int_{r_0}^{r_*} \frac{G P}{\sqrt{P^2 - P_0^2}} dr - \int_{r_H}^{r_*} G dr \right] \ , \label{potqq} \\
&& {\rm with} \quad P = e^{\Phi/2} \sqrt{G_{tt} G_{xx}} \ , \quad G = e^{\Phi/2} \sqrt{G_{tt} G_{rr}} \ .
\end{eqnarray}
Once again, $G$ above represents the background metric and the dilaton enters the above formulae since we are working in the Einstein frame. Our task here is to study the potential $V$ as a function of the quark--anti-quark separation $\ell$.

\begin{figure}[h!]
\centering
\includegraphics[scale=0.65]{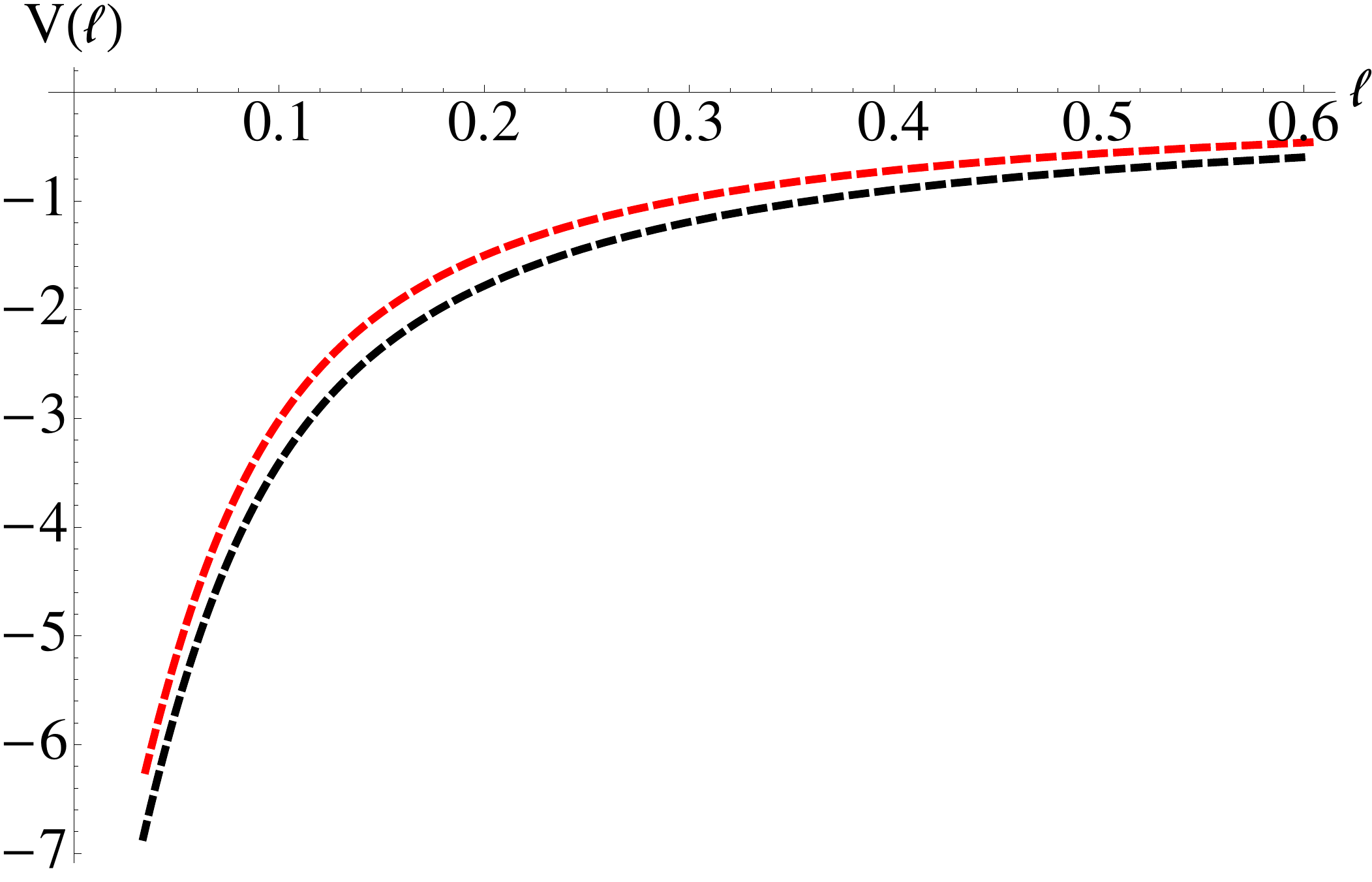}
\caption{The quark--anti-quark potential obtained from (\ref{lenqq}) and (\ref{potqq}). The black dashed line corresponds to $\epsilon =0$ and the red dashed line correspond to $\epsilon =0.5$. We have further set $L=1$, $r_H \ll 1$, $\alpha = 0$. }
\label{qqpot}
\end{figure}

Recall that at zero temperature, in the absence of any back-reaction, the quark--anti-quark potential is Coulombic \cite{Maldacena:1998im}; at finite temperature it undergoes a phase transition depending on the value of $T\ell$, as observed in \cite{Brandhuber:1998bs, Rey:1998bq}. In the presence of the back-reaction, this phase transition will continue to exist and will receive $\epsilon$ order corrections. Thus we do not expect any qualitative change in the thermal physics of the system. Therefore we will consider the case when $T\ell \ll 1$ and explore how the Coulomb potential is affected by the back-reaction.\footnote{For some related studies in similar models, see {\it e.g.} \cite{Bigazzi:2008ie}.} A representative plot is shown in fig.~\ref{qqpot}.
It is evident that the Coulomb behaviour of the potential does not change in the presence of the back-reaction. Although we have not explicitly presented it, this behaviour does not change when $\alpha \not = 0$. This is in accord with what has been observed in {\it e.g.} \cite{Bigazzi:2008ie}.

\subsection{Entanglement entropy}

Entanglement entropy is a measure of quantum entanglement of a given system. It is defined as the von Neumann entropy of a reduced density matrix. For a given system, let us imagine dividing it into two parts denoted by $A$ and $B$. For an observer who is restricted to access the information of the subsystem $A$ only, the system will be described by the reduced density matrix $\rho_A = {\rm tr}_B \rho_{\rm tot}$, where $\rho_{\rm tot}$ denotes the total density matrix of the full system. Now, the entanglement entropy of the subsystem $A$ is defined as 
\begin{eqnarray}
S_A = - {\rm tr} _A \rho_A \log \rho_A \ .
\end{eqnarray}
In the context of AdS/CFT correspondence, a proposal to compute the entanglement entropy was suggested in \cite{Ryu:2006bv, Ryu:2006ef}. Suppose we divide the system into two regions: one ``rectangular" strip of length $\ell$ along $x^1$-direction (denoted by $A$) and its complement. The ``rectangular" strip obviously has an infinite dimension along the $x^2$ and the $x^3$-directions. In such a scenario, the entanglement entropy is obtained by computing the minimal area surface whose boundary coincides with the boundary of the region $A$.

In the $10$-dimensional Einstein frame metric, the Ryu-Takayanagi proposal yields 
\begin{eqnarray} \label{ee}
S_A = \frac{1}{4 G_{10}} \int d^8\xi \sqrt{{\rm det} G}  = \frac{2}{27 G_{10}} V_{\mathbb R^2}\int \sqrt{h} e^{f+4g} \left( x'^2 + \frac{h}{b} \right)^{1/2} \ ,
\end{eqnarray}
where $V_{\mathbb R^2}$ denotes the area of the ``rectangular" strip along the $\{x^2, x^3\}$-direction. The minimal area surface is parametrized by $x^1(r) \equiv x(r)$ with the boundary conditions $x(r_*) = \pm \ell/2$. Before going further, let us note that for the perturbative solution presented in (\ref{solb})-(\ref{solg}) the quantity $(f + 4g)$ is independent of the coupling $\mathfrak{m}$. Hence the dimension $6$ operator will have vanishing contribution to the entanglement entropy at least at the leading order in $\epsilon$. This is not true for the dimension $8$ operator, whose coupling is denoted by $\alpha$.

The equation of motion resulting from minimizing the volume functional is given by
\begin{eqnarray}
\sqrt{h} e^{f+4g} \frac{x'}{\sqrt{x'^2 + \frac{h}{b}}} = {\rm const} = \sqrt{h_0} e^{f_0 + 4 g_0} \ ,
\end{eqnarray}
where we have imposed the condition that at $r= r_0$ the minimal area surface turns over and have defined: $h_0 = h(r_0)$, $f_0 = f(r_0)$ and $g_0 = g(r_0)$. Substituting back the profile of the minimal area surface in (\ref{ee}), we get
\begin{eqnarray}
S_A = S_A^{(0)} + \epsilon S_A^{(1)} \ ,
\end{eqnarray}
at the leading order in $\epsilon$.

One crucial property of the entanglement entropy is the so called ``area law", {\it i.e.} $S_A$ scales as the area of the sub-region $A$: $S_A \sim (\partial A)/a^{2}$, where $a$ is an infrared cut-off in the dual field theory. It is easy to check that $S_A^{(0)} \sim (\partial A)/a^{2}$ and there are no sub-leading divergent pieces, as was previously alluded to in \cite{Ryu:2006ef};\footnote{For the finite part of the entanglement entropy and physics related to it see \cite{Fischler:2012ca, Fischler:2012uv}.} the new term here is $S_A^{(1)}$. Using the explicit functions for the background in (\ref{solb})-(\ref{solg}), it is straightforward to check that
\begin{eqnarray}
S_A^{(1)} \sim \left( \partial A\right) \frac{r_*^2}{24}  \left( 8 r_*^4 \alpha - L^4 \right) + {\rm finite} \ ,
\end{eqnarray}
where $r_*$ is the UV cut-off and we can identify $a^{-1}\equiv r_*$. Note that the term $\epsilon \alpha r_*^4$ has to be small because of the condition in (\ref{cond2}). Hence, the divergence structure is the same as in the case where the back-reaction vanishes. It is nonetheless an interesting question whether going beyond a leading order perturbative solution in $\epsilon$ changes the divergence structure of the entanglement entropy.

\section{Physics on the probe brane}

We will investigate the effects of introducing a constant electromagnetic field on the additional probe sector in this model. For all figures and plots exploring the various phase structures will assume $\alpha=0$ henceforth. Let us begin by discussing the case of a purely electric field.

\subsection{Purely electric field}

Let us introduce a gauge field in the additional probe sector of the form
\begin{eqnarray}
A_{x^1} = -E t + A_1(r) \ ,
\end{eqnarray}
where $E$ is the electric field along the $x^1$-direction. The function $A_1(r)$ encodes the possibility of a non-zero current which results from applying the electric field. The Euclideanized DBI action is given by
\begin{eqnarray}
S_{\rm DBI} & = & \cN_T \int dr e^{f+2g+\Phi} \left[ \left(1 - \frac{e^2 h}{b}\right) \left(1 + \frac{b}{6} e^{2g} \phi'^2 \right) + b a_1'^2 \right]^{1/2} \ , \\
e & = & \left(2 \pi\alpha'\right) E \ , \quad a_1 = \left(2 \pi\alpha'\right) A_x \ .
\end{eqnarray}
As discussed in \cite{Alam:2012fw}, for a non-trivial $\phi(r)$, the action is minimized for $a_1' = 0$. To have a non-trivial $a_1'$, we focus on the parallel-shaped profiles which are given by $\phi = {\rm const}$. For the parallel-shaped profiles the equation of motion for the gauge field is given by
\begin{eqnarray}\label{gaugesol0}
&& e^{f+2g + \Phi} \frac{b a_1'}{\left( 1 - \frac{e^2 h}{b} + b a_1'^2 \right)^{1/2}} = j \ , \nonumber\\
&& \implies a_1' = \frac{j}{b} \left(\frac{ b - e^2 h}{ b e^{2f+4g+2 \Phi} - j^2 }\right)^{1/2} \ . \label{gaugesol}
\end{eqnarray}
Asymptotically, the solution for the gauge field takes the following form
\begin{eqnarray}
a_1(r) = - \frac{j}{2 r^2} \left( 1 + \epsilon \frac{1 - 32 \alpha e^2}{32} \right) + \epsilon \left(3 j \mathfrak{m}^{-2}\right) \log r + \ldots \ ,
\end{eqnarray}
The condition that the solution for the gauge field given in equation (\ref{gaugesol0}) remain real ultimately determines the constant $j$ in terms of the electric field and other parameters of the theory.

First, the location of the pseudo-horizon --- which is where the numerator of the right hand side of (\ref{gaugesol0}) vanishes --- is given by
\begin{eqnarray} \label{phE}
r_{\rm ph} ^4 & = & \left(e^2 L^4 + r_H^4 \right)  + \frac{\epsilon}{8} e^2 \left[ 8 r_H^4 \alpha + L^4 \left( 1 + 16 e^2 \alpha \right) \right] + \cO(\epsilon^2) \ .
\end{eqnarray}
Note that in the above formula the coupling $\mathfrak m$ does not enter. In general, depending on the sign of the constant $\alpha$, the location of the pseudo-horizon can increase or decrease compared to the location in the case of vanishing back-reaction. If we set $\alpha = 0$, then the pseudo-horizon receives a positive contribution coming from the back-reaction. Now the constant $j$ can be fixed by demanding 
\begin{eqnarray}\label{jegen}
j (\alpha, \mathfrak{m}, e, r_H) = \left. b(r)^{1/2} e^{f(r) + 2g(r) + \Phi(r)} \right|_{r=r_{\rm ph}} \ .
\end{eqnarray}
In general the correction to the constant $j$ at the leading order in $\epsilon$ has a complicated algebraic form. For illustrative purposes, we can present one simplifying case in the limit $r_H^2 / (e L^2) \ll 1$. We get
\begin{eqnarray}
j^2 = \left(e L^2\right)^{3} \left[ 1 + \frac{\epsilon}{48} - 8 \epsilon \left( e L^2 \right) \mathfrak{m}^{-2} + 3 \epsilon e^2 \alpha + \frac{\epsilon}{8} \log \left( \frac{r_{\rm ph}}{r_*} \right) \right] + \cO\left(\frac{r_H}{\sqrt{e}L} \right) \ ,
\end{eqnarray}
where we have used equation (\ref{phE}) above.

To recast the above formula in terms of the quantities defined on the boundary field theory, let us recall a few basic definitions: First, the overall constant $\cN$ that appears in front of the DBI action leaving out the integral over the time direction
\begin{eqnarray}
\cN = N_f' \tau_7 \left( 8 \pi^2 \right) \ , \quad {\rm and} \quad \tau_7 = \frac{1}{g_s^*} \left(2 \pi\right)^{-7} \alpha'^{-4} \ ,
\end{eqnarray}
where we have set the volume of the three Minkowski spatial directions, $V_{\mathbb{R}^3} =1$. If we also set the radius of the deformed AdS-space, $L=1$ then using equation (\ref{radius}) and the above definitions we get
\begin{eqnarray}
\cN = \overline{\lambda} N_f' N_c \ , \quad {\rm where} \quad \overline{\lambda} = \frac{3^4}{2^{6}} \lambda = \frac{3^4}{2^{6}} \left(4\pi g_s^* N_c \right) \ ,
\end{eqnarray}
where $\lambda$ denotes the 't Hooft coupling. The current density in the boundary gauge theory, denoted by $\langle J_x \rangle$, can be obtained as (see equation (\ref{current}))
\begin{eqnarray}
\langle J_x \rangle = \cN \left(2 \pi \alpha' \right) j \ .
\end{eqnarray}
For simplicity, setting $\cm^{-1} = 0 = \alpha$ we get
\begin{eqnarray}
\langle J_x \rangle = \overline{\lambda} N_f' N_c E \sqrt{ 2 \pi \alpha' E} \left( 1 + \frac{\epsilon}{96} + \frac{\epsilon}{16} \log\left(\frac{r_{\rm ph}}{r_*}\right)\right) \ .
\end{eqnarray}
Here we can define an effective 't Hooft coupling absorbing all $\epsilon$ dependences in it, which will leave us with a simple formula for the conductivity much like for the case of vanishing back-reaction.

\begin{figure}[h!]
\centering
\includegraphics[scale=0.78]{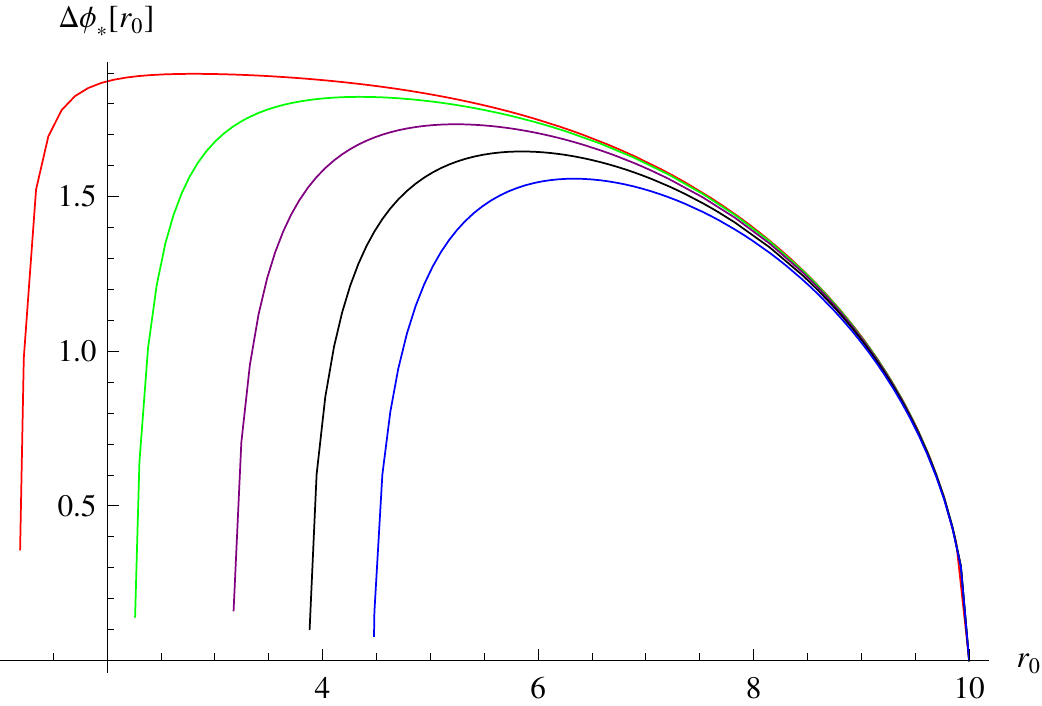}
\caption{$\cphi(r_0)$ for $r_H=1, \epsilon=0.01, m^2=0.05$, $\alpha=0$ and $e=1$ (red), $e=5$ (green), $e=10$ (purple), $e=15$ (black) and $e=20$ (blue). Clearly $e$ is measured in units of $r_H^2$.}
\label{fig71}
\end{figure}

\begin{figure}[h!]
\centering
\includegraphics[scale=0.9]{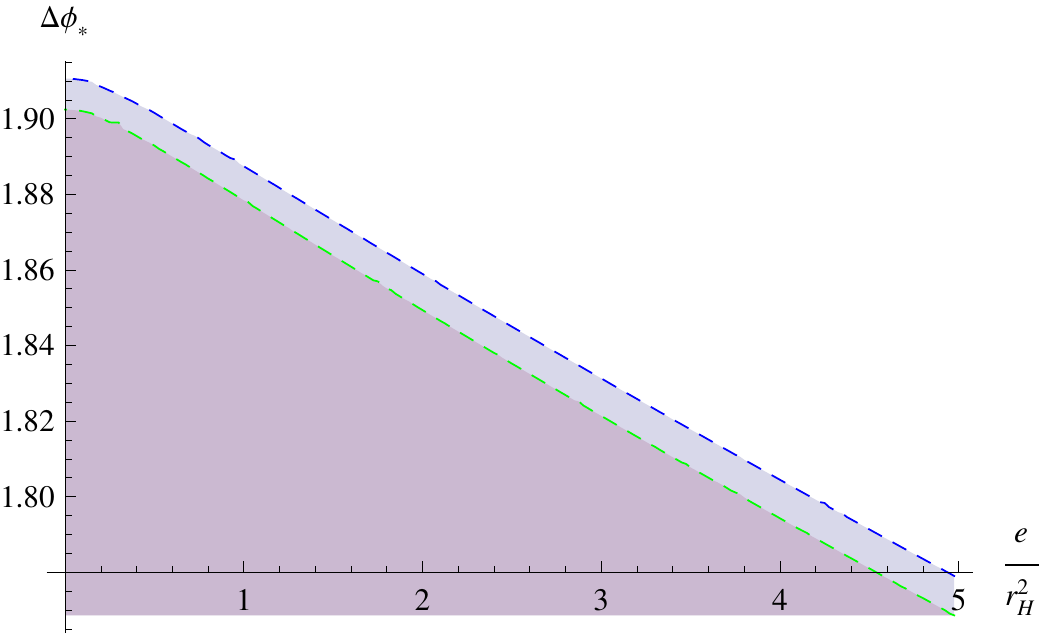}
\caption{The critical $\cphi \left(\frac{e}{r_H^2} \right)$ for $m^2=0.001$ and $\epsilon=0.01$ (blue) or $\epsilon=0.1$ (green). The shaded region below the respective dashed lines corresponds to the $\chi$SB or metastable $\chi$SR phase and the region above the respective dashed lines represents
the $\chi$SR or metastable $\chi$SB phase.}
\label{fig8}
\end{figure}

Notice that now, even for the purely electric field case, we will have a non-trivial phase structure. To demonstrate the existence of this phase transition, in fig.~\ref{fig71} we have shown the behaviour of $\cphi$ as a function of  $r_0$, the radial position where the brane--anti-brane pair joins, for different values of $e/r_H^2$.  
On the other hand our expectation is that increasing the electric field will favour a restoration of the chiral symmetry. Thus we should observe a monotonically decreasing behaviour of the phase boundary curve in the $\cphi$ vs $e/r_H^2$ plane. This is demonstrated in fig.~\ref{fig8}. We should note that only for $\epsilon \not =0$ there is a non-trivial phase structure and the qualitative features are similar for various values of $\epsilon$. 
%

\subsection{Purely magnetic field}

Now we will introduce a constant magnetic field on the worldvolume of the additional probe D7 and anti-D7 branes. The ansatz for the gauge field is 
\begin{eqnarray}
A_3 = H x^2 \ , \label{magfield}
\end{eqnarray}
which represents a constant magnetic field $F_{23} = H $ along the $x^1$-direction. With this gauge field the DBI action is given by:
\begin{equation}
S_{\rm DBI}= \cN_T  \int dr e^{f(r)+2 g(r)+\Phi (r)} \left(1+4 \pi ^2 \alpha'^2  H^2 h(r)\right)^{1/2}\sqrt{1+\frac{1}{6} b(r) e^{2 g(r)} \phi '(r)^2} \ .
\end{equation}
Defining $B:= 2 \pi \alpha' H $, we obtain
\begin{equation}
\frac{e^{f+4g+\Phi}\sqrt{1+B^2 h(r)}\frac{b}{6}\phi'}{\sqrt{1+\frac{1}{6} b e^{2g} \phi'^2}} = c_H \ . 
\end{equation}
Equivalently, 
\begin{equation}
\phi '(r)=\frac{6 c_H}{\sqrt{b(r)^2 \left(1+B^2 h(r)\right) e^{2 f(r)+8 g(r)+2 \Phi (r)} - 6 c_H^2 b(r) e^{2 g(r)}}} \ ,
\end{equation}
where we have defined 
\begin{eqnarray}
c_H^2 = \frac{e^{2f(r_0) + 6g(r_0) + 2\Phi(r_0)}}{6} b(r_0) \left(1+ B^2 h(r_0)\right) \ .
\end{eqnarray}

\begin{figure}[h!]
\centering
\includegraphics[scale=0.78]{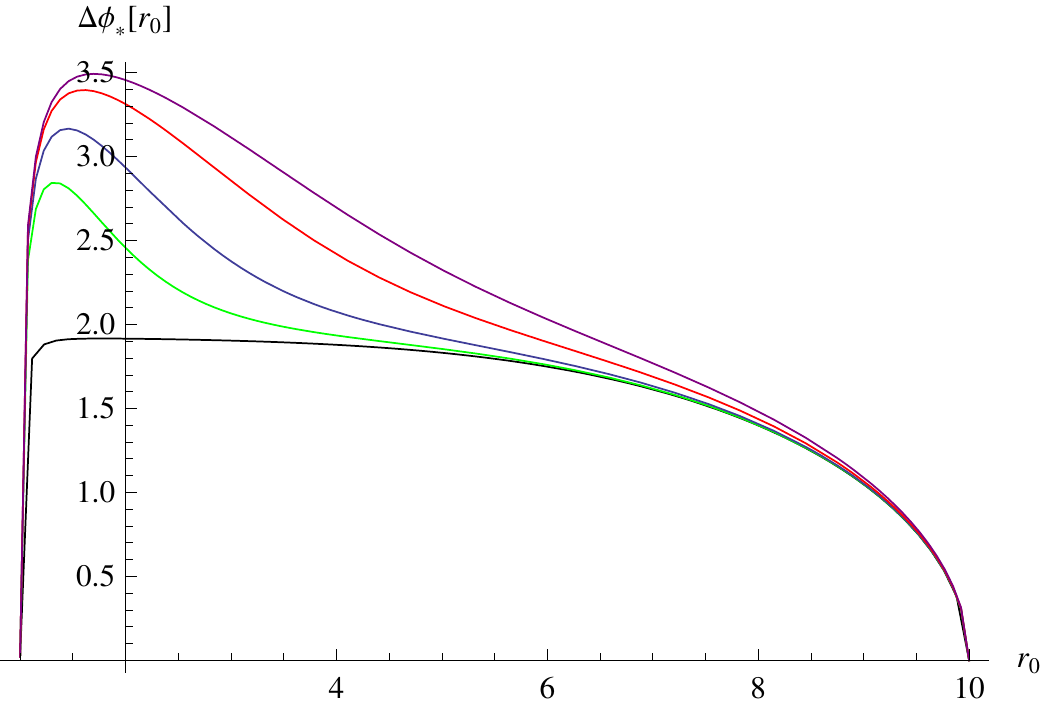}
\caption{$\cphi(r_0)$ for $r_H=1, \epsilon=0.01, m^2=0.05$, $\alpha=0$ and $B=0$ (black), $B=5$ (green), $B=10$ (blue), $B=20$ (red) and $B=30$ (purple). }
\label{fig5}
\end{figure}

\begin{figure}[h!]
\centering
\includegraphics[scale=0.9]{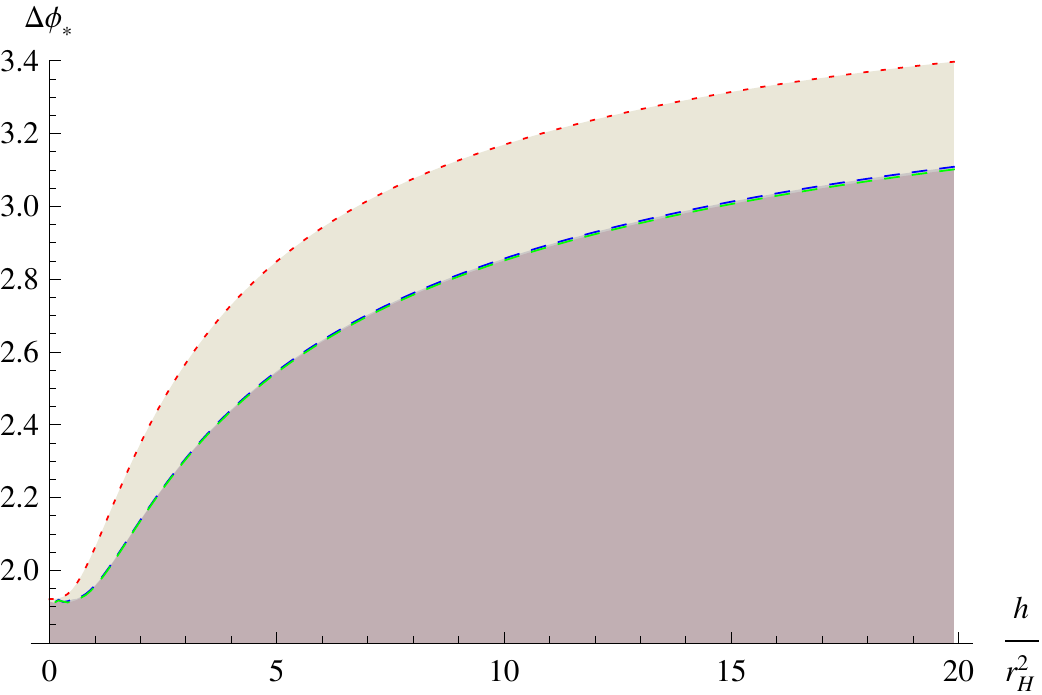}
\caption{The critical (blue and green, dashed) and maximal (red, dotted) $\Delta \phi_{\ast}(h/r_H^2)$ for $m^2=0.001$ and $\epsilon=0.01$ (blue, red) or $\epsilon=0.05$ (green). The shaded region below the dashed line corresponds to the $\chi$SB or metastable $\chi$SR phase and the region between the dashed and dotted lines represents
the $\chi$SR or metastable $\chi$SB phase. Above the dotted line, $\chi$SR is the only possible configuration.} 
\label{fig6}
\end{figure}

To demonstrate how this constant magnetic field affects the coupling $\cphi$, we can make a plot in the $\cphi$ vs $r_0$ plane for different values of $B$. This is shown in fig.~\ref{fig5}. The existence of a phase transition is self-evident from this diagram and the corresponding phase structure is shown in fig.~\ref{fig6}, which is consistent with the phenomenon of magnetic catalysis. 
%

\subsection{Holographic renormalization}

In the context of the AdS/CFT correspondence, the on-shell action $S_{\rm DBI }$ of the additional probe sector corresponds to the generating functional of the additional flavour sector that has been introduced in the dual field theory. This on-shell action contains UV divergences. Holographic renormalization is the rigorous procedure to regulate such divergences by adding covariant counter-terms on a cut-off surface and then taking the cut-off to infinity. In our current scenario, we have a cut-off surface $r_*$ already in the problem and we need to restrict this cut-off surface to a position much below the Landau pole, which in our case is located at infinity. For a review on the procedure of holographic renormalization, see {\it e.g.} \cite{Skenderis:2002wp}.

Let us apply the procedure of holographic renormalization to our current case. Here we will work with Euclidean signature. To procede, we need the following data: $\Lambda \equiv$ cut-off; $\gamma_{ij}$ is the induced metric on the $r=\Lambda$ slice and $\gamma : = {\rm det} \gamma_{ij}$. After the counter-terms are introduced, we need to take $\Lambda \to r_*$, which is the actual cut-off surface.

Firstly, we shall discuss the case of the parallel embeddings. We will review and elaborate on the discussion presented in \cite{Ihl:2012bm}. In the absence of any external fields, the on-shell Euclidean action for the parallel embeddings is given by
\begin{eqnarray} \label{osp}
S_{||} = \cN_T \int_{r_H}^{r_*} dr e^{f(r) + 2 g(r) + \Phi(r)} \ .
\end{eqnarray}

The divergent pieces in $S_{||}$ can be arranged as follows:
\begin{eqnarray}
S_{||}^{\rm div} = \cN_T \left[ - \frac{2 r_*^6 \left(\cm^{-2} \epsilon\right)}{3} + \frac{1}{192} r_*^4 \left( - 7 \epsilon + 48 \right) + \frac{1}{2} \cm^{-2} r_*^2 \epsilon r_H^4 - \frac{1}{16} \epsilon r_H^4 \log\left(\frac{r_H}{r_*}\right)\right] + {\rm finite} \ .
\end{eqnarray}
Of course, the $\epsilon$ dependent terms are parametrically much smaller compared to order one numbers and thus the leading order divergent behaviour is identical to the pure AdS-case, which contains only a quartic term. In addition to that there is another divergent piece proportional to $(\epsilon r_*^6)$. The remaining terms in the above expression are actually not divergent under the conditions written in equations (\ref{cond1}) and (\ref{cond2}) and will eventually be absorbed in an effective 't Hooft coupling which receives correction due to the presence of the back-reaction and the temperature. It is worthwhile to remark that the ``potentially divergent" term proportional to $\log(r_H/r_*)$ stems from the infrared part of the geometry. This can be understood by noticing that this term will not arise if one first expands the integrand in (\ref{osp}) in inverse powers of $r_*$ and then integrates; instead one needs to first integrate (\ref{osp}) and then expand the result in inverse powers of $r_*$.  Note that the on-shell action does not depend on the warp factor $h(r)$ and hence the divergences are also insensitive to the constant $\alpha$.

It was shown in \cite{Ihl:2012bm} that we need only one counter-term to regulate the divergences of this on-shell action; however, the $\alpha$-dependent contribution (which will be present in the counter-term) was not included there. In the most general case, we need the following counter-term
\begin{eqnarray} \label{countn}
S_{\rm ct} & = & - \cN_T \frac{L}{4} \left(1 - \epsilon \frac{1}{32}  + \frac{4}{3} \epsilon r_*^2 \cm^{-2} + \frac{\alpha}{2 L^4} \epsilon r_*^4 \right) \sqrt{\det \gamma} \nonumber\\
& = & \cA\left( \epsilon \cm^{-2} r_*^2 , \epsilon \alpha r_*^4\right) \sqrt{\det \gamma} \ .
\end{eqnarray}
The constant $\cA$ receives {\it finite} correction coming from the presence of the back-reaction and the irrelevant operators in the theory. This leaves us with the two ``potentially divergent" pieces which can be absorbed in the definition of an effective 't Hooft coupling: 
\begin{eqnarray}
\lambda_{\rm eff} (T) = \lambda \left[ 1 + \frac{1}{4} \epsilon \log \left(\frac{r_H}{r_*}\right) + \frac{1}{3} \epsilon \cm^{-2} r_*^2 \right] + \cO(\epsilon^2) \ ,
\end{eqnarray}
where $\lambda = 4 \pi g_s^* N_c$ is the 't Hooft coupling. Note that the effective 't Hooft coupling also receives finite corrections due to the presence of the back-reaction, the deformations of the original CFT and the background temperature. This exercise demonstrates that the presence of the back-reaction does not call for a new counter-term, at least at the leading order in $\epsilon$. Let us now discuss the case when there is a constant electro-magnetic field on the worldvolume of the probe sector. We will discuss the electric and the magnetic cases separately. \\

\noindent{\bf Case I: electric field}\\

We will first discuss some subtleties that occur in the presence of an electric field as elaborated in \cite{Alam:2012fw}. First, the on-shell action in this case for the parallel embeddings should be supplemented by a boundary term due to the variation of the gauge field itself. Second, the lower limit of the integration is where the pseudo-horizon is located rather than the actual event-horizon of the background. The latter is motivated by a couple of facts, most prominently the fact that the open string metric, which is the metric the open string degrees of freedom should sense, possesses an event-horizon which is the pseudo-horizon; and hence it is natural to cut-off the integral at this location. This prescription also has the technical advantage of avoiding an IR singularity coming from the location of the event-horizon \cite{Alam:2012fw}. Altogether, we arrive at the following form for the on-shell action,
\begin{eqnarray} \label{osE}
S_{||} = \cN_T \int_{r_{\rm ph}}^{r_*} dr \left[ e^{2f(r) + 4g(r) + 2 \Phi(r)} \frac{\left(b(r) - e^2 h(r)\right)^{1/2}}{\left(b(r) e^{2f(r)+4g(r)+2\Phi(r)} - j^2 \right)^{1/2}} - j a_1' \right] \ ,
\end{eqnarray}
where $r_{\rm ph}$ was determined in equation (\ref{phE}). Subsequently the constant $j$ is determined from equation (\ref{jegen}) and the gauge field can be obtained from equation (\ref{gaugesol0}).

It is straightforward to see that the boundary term due to the gauge field contributes a finite quantity and there will be no divergences associated with it. All divergences will come from the DBI-piece. Let us comment on the case of vanishing back-reaction. The presence of the background gauge field introduces a new logarithmic divergence,
\begin{eqnarray}
S_{||}^{\rm div} = - \cN_T \left(\frac{r_*^4}{4} - \frac{1}{2} e^2 L^4 \log r_* \right) + {\rm finite} \ ,
\end{eqnarray}
where $r_*$ is the cut-off surface. The above divergences can be regulated by adding a counter-term on the $r={\rm const}$. slice
\begin{eqnarray}
S_{\rm ct} & = & \cN_T \frac{L}{4} \left( \sqrt{{\rm det} \gamma} - \left(2 \pi \alpha' \right)^2 \sqrt{{\rm det} \gamma} \, \gamma^{ij} \gamma^{kl} F_{ik} F_{jl} \log r_* \right)  \nonumber\\
& = & \cA \sqrt{{\rm det} \gamma} + \cB \sqrt{{\rm det} \gamma} \, \gamma^{ij} \gamma^{kl} F_{ik} F_{jl} \log r_* \ , 
\end{eqnarray}
where $\cA$ and $\cB$ are the coefficients of the counter-terms.

Now --- including the effect of the back-reaction --- the divergences take the following form:
\begin{eqnarray} \label{divE}
S_{||}^{\rm div} / \cN_T & = & \frac{r_*^4}{4} \left( 1 - \frac{7 \epsilon}{48} \right) - \frac{1}{2} e^2 L^4 \log r_* \left(1 + \frac{\epsilon}{24} \right) \nonumber\\
& + & \epsilon \left[ - \frac{2}{3}\cm^{-2} r_*^6 + \frac{1}{2} \cm^{-2} r_*^2 \left(r_{\rm ph}^4 +e^2 L^4 \right) - \frac{1}{4} e^2 \alpha r_*^4 + \left( 2 j^2 \cm^{-2} - \frac{1}{2} \alpha e^2 r_{\rm ph}^4 \right) \log \left(\frac{r_*}{r_{\rm ph}}\right) \right] \nonumber\\
& + & \frac{\epsilon}{16} e^2 L^4 \left(\log (r_*/r_{\rm ph})\right)^2 +  \cJ(r_{\rm ph}, e, r_*) + {\rm finite} \ , 
\end{eqnarray}
which is true at the leading order in $\epsilon$. The term $\cJ$ is a ``potentially divergent" term coming from the IR of the geometry, which in this case is located at $r= r_{\rm ph}$. This term gives
\begin{eqnarray}
\cJ \left(r_{\rm ph}, e, r_*\right) = \frac{\epsilon}{2} \int_{r_{\rm ph}} \frac{x^5}{\sqrt{4 x^4 + 2 e^2 L^4}} \log\left(\frac{x}{r_*}\right) dx  \ ,
\end{eqnarray}
which can be integrated analytically; however, we refrain from doing so since the result is algebraically complicated and not particularly illuminating. Note that, when $e \to 0$, we get
\begin{eqnarray}
\cJ(r_{\rm ph}, 0, r_*) = \cJ(r_H, 0, r_*) = - \frac{\epsilon}{16} r_H^4 \log\left(\frac{r_H}{r_*}\right) \ ,
\end{eqnarray}
which is exactly what we had in \cite{Ihl:2012bm}. Ultimately, the term $\cJ$ contributes to an effective 't Hooft coupling as explicitly demonstrated in \cite{Ihl:2012bm}.

We will now introduce appropriate counter-terms to take care of the divergences. Note that the leading order divergences in (\ref{divE}) are again a quartic one and a logarithmic one. There is an $r_*^6$ divergence supported purely by the back-reaction. The rest of the terms --- which depend on $r_*$ --- are nonetheless parametrically finite within the regime of validity of our solution. Since the physics should not depend on the cut-off surface $r_*$, we will define an effective coupling which will receive corrections due to the presence of the horizon, the electric field and the cut-off surface.

The counter-term turns out to be:
\begin{eqnarray} \label{ctE}
S_{\rm ct} & = & - \cN_T \frac{L}{4} \left[ \left( 1 - \frac{\epsilon}{32} + \frac{\epsilon}{2L^4} \alpha r_*^4 + \frac{4\epsilon}{3} \cm^{-2} r_*^2 + 3 \alpha e^2 \epsilon \log(r_*/r_{\rm ph}) \sqrt{{\rm det} \gamma}\right) \right. \nonumber\\
&& \left.  -  \left(1 + 4\epsilon \cm^{-2} r_*^2 - \frac{\epsilon}{8} \log(r_*/r_{\rm ph})\right) \left(2 \pi \alpha' \right)^2 \sqrt{{\rm det} \gamma} \, \gamma^{ij} \gamma^{kl} F_{ik}F_{jl}  \log(r_*/ r_{\rm ph}) \right] \nonumber\\
& = & \cA \sqrt{{\rm det} \gamma} + \cB \sqrt{{\rm det} \gamma} \, \gamma^{ij} \gamma^{kl} F_{ik} F_{jl} \log r_* \ .
\end{eqnarray}
which again implies that there is no need for any additional counter-term as compared to the case of vanishing back-reaction. The presence of the back-reaction and other relevant parameters in the theory, such as the temperature or the electric field, yields a finite contribution to the coefficients of the counter-terms.

To obtain the boundary current, one can follow the procedure outlined in \cite{Karch:2007pd}. We will briefly review this process and argue that the presence of the back-reaction does not change the identification of the boundary current. To this end, we go back to Minkowski signature and write the on-shell action for the probes as
\begin{eqnarray}
S_{D7} = - \cN \int dt dr \cL_{\rm on-shell} + \cN \int dt \cL_{\rm ct} \ ,
\end{eqnarray}
where $\cL_{\rm on-shell}$ denotes the on-shell Lagrangian (presented in equation (\ref{osE})) and $\cL_{\rm ct}$ denotes the counter-terms given in equation (\ref{ctE}) and $\cN = T \cN_T$, where $T$ is the background temperature.

The variation of the on-shell regularized action is
\begin{eqnarray}
\delta S_{D7} = - \cN \left[ \int dt dr \left(\frac{\delta \cL_{\rm on-shell}}{\delta \partial_r A_x} \partial_r \delta A_x + \frac{\delta \cL_{\rm on-shell}}{\delta \partial_t A_x} \partial_t \delta A_x \right)  - \int dt \frac{\delta \cL_{\rm ct}}{\delta \partial_t A_x} \partial_t \delta A_x \right] \ .
\end{eqnarray}
There will be no contribution coming from the counter-term since we impose $\int dt \partial_t \delta A_x = 0$. The only contribution will come from the first variation of the on-shell Lagrangian and thus we get
\begin{eqnarray} \label{current}
\langle J_x \rangle:= \frac{\delta S_{D7}}{\delta A_x} = \cN \left(2 \pi \alpha' \right) j \ ,
\end{eqnarray}
where we have used the definition of $j$ from equation (\ref{gaugesol}). \\

\noindent{\bf Case II: magnetic field}\\

Since the divergence structure is identical for the parallel and the U-shaped embeddings, we will discuss the parallel case in detail. The on-shell action with a purely magnetic field is given by
\begin{eqnarray}
S_{||} = \cN_T \int dr e^{f(r) + 2 g(r) + \Phi(r)} \left(1 + B^2 h(r) \right)^{1/2} \ .
\end{eqnarray}
The divergences of the on-shell action takes the following form
\begin{eqnarray} \label{divB}
S_{||}/ \cN_T & = & \frac{r_*^4}{4} \left( 1 - \frac{7\epsilon}{48}\right) + \frac{1}{2} B^2 L^4 \log r_* \left(1 + \frac{\epsilon}{24} \right) \nonumber\\
& + & \epsilon \left[ - \frac{2}{3} \cm^{-2} r_*^6 + \frac{1}{4} B^2 \alpha r_*^4 + \frac{1}{2} \cm^{-2} r_*^2 \left(r_H^4 - 2 B^2L^4\right) - \frac{1}{2} \alpha r_H^4 B^2  \log \left(\frac{r_*}{r_H}\right) \right. \nonumber\\
 & - & \left. \frac{1}{16} B^2 L^4 \left( \log(r_*/r_{H}) \right)^2 \right] + \cJ \left( r_*, B, r_H\right) + {\rm finite} \ ,
\end{eqnarray}
where $\cJ$ is the ``potentially divergent" term that arises from the IR of the background, which is explicitly given by
\begin{eqnarray}
\cJ = \frac{\epsilon}{4} \int_{r_H} dx  x \sqrt{x^4 + B^2L^4} \log\left(\frac{x}{r_*}\right) \ .
\end{eqnarray}
Note that the divergences in equation (\ref{divB}) are very similar to the ones encountered for purely electric case in equation (\ref{divE}); in fact, equation (\ref{divB}) can be obtained from equation (\ref{divE}) by $j \to 0$ and $e^2 \to - B^2$, which of course makes sense since in the absence of a finite temperature Lorentz symmetry allows one to the identify the physics by sending $e^2 \to - B^2$. This means that the divergence structure does not depend on the finite temperature. However, the presence of the finite temperature affects the effective 't Hooft coupling in inequivalent ways for the electric and the magnetic cases.

\subsection{Electromagnetic fields}

At zero temperature, in the presence of an electric and magnetic field, denoted by $E$ and $H$ respectively, there are two Lorentz invariants: $\vec{E}^2 - \vec{H}^2$ and $\vec{E} \cdot \vec{H}$. Therefore, when the temperature vanishes, it is sufficient to analyze the cases when the electric and the magnetic fields are parallel or perpendicular. However, it was demonstrated in \cite{Alam:2012fw} that these two situations yield qualitatively similar results; hence in order to keep our discussions simple, we will consider the perpendicular case only.

Let us introduce 
\begin{equation}
 A_{x^1} = -E t + A_1(r) \ , \quad A_{x^2} = H x^1  \ ,
\end{equation}
which yields
\begin{align*}
S_{\rm DBI} & = \cN_T \int dr e^{f(r)+2 g(r)+\Phi (r)} \left[\left(1+ B^2 h(r) - e^2 \frac{h(r)}{b(r)}\right)\right.\cr 
& \qquad \qquad \left. \left(1+\frac{1}{6} b(r) e^{2 g(r)} \phi '(r)^2 \right) + b(r) \left(\frac{\partial a_1(r)}{\partial r}\right)^2 \right]^{1/2} \ .
\end{align*}
For convenience, we have again introduced the following:
\begin{equation}
e:= 2 \pi \alpha' E \ , \quad B:= 2 \pi \alpha' H \ , \quad a_1:= 2 \pi \alpha' A_1 \ . 
\end{equation}
Of course, the action should be supplemented by the boundary term discussed in \cite{Alam:2012fw}, but we have not written it explicitly since it does not affect the equations of motion. The boundary term becomes relevant only for the computation of the free energy and the discussion of the phase diagrams. The basic structure of the probe profile functions remain qualitatively the same as discussed in \cite{Alam:2012fw}. In the case of a non-trivial profile $\phi(r)$, the action is clearly minimized when $a_1'(r) =0$.

For the parallel embeddings we find, 
\begin{align} \label{gaugesol}
a' & = \pm \frac{j}{ \sqrt{b( r )}} \frac{\sqrt{1 + B^2 h(r) - e^2 \frac{h(r)}{b(r)}}}{\sqrt{b(r ) e^{2f(r ) +4g( r) +2 \Phi (r )} -j^2}} \ ,\\
    & = \pm \left[ \frac{j}{r^3} + \mathcal{O}(r^{-7})+ \epsilon \left( \frac{4j \mathfrak{m}^{-2}}{r} + \mathcal{O}(r^{-3})\right)\right] + \ldots \ .
\end{align}
The second line above is valid as $r \to r_*$, the UV cut-off. As it was pointed out in \cite{Alam:2012fw}, imposing the ingoing boundary condition singles out the solution in (\ref{gaugesol}) with positive sign.

The location of the the pseudo-horizon, denoted by $r_{\rm ph}$, is now obtained by solving the following algebraic equation
\begin{equation}
\left. 1 + B^2 h(r) - e^2 \frac{h(r)}{b(r)} \right|_{r = r_{\rm ph}} = 0 \ ,
\end{equation}
which then fixes the response-current 
\begin{eqnarray}
j = \left. b(r) e^{2 f(r) + 4 g(r) + 2 \Phi(r)} \right|_{r=r_{\rm ph}} \ .
\end{eqnarray}
The current in the dual field theory is proportional to this constant $j$, with the proportionality constant determined in (\ref{current}). It is not particularly illuminating to present the formulae explicitly in terms of all the variables of the system; therefore we refrain from doing so. The formula above gives an $N_f/N_c$ correction of the conductivity formula discussed in \cite{Karch:2007pd, Albash:2007bq}. Note that in \cite{Magana:2012kh} an analogous $N_f/N_c$ correction to the conductivity formula has been obtained for a probe sector which is different from what we are discussing here. So, although in the limit of vanishing back-reaction, both our result and the result of \cite{Magana:2012kh} coincide with \cite{Karch:2007pd}, the $N_f/N_c$ correction is different.

As far as the possibility of a phase transition is considered, let us investigate (along the lines of \cite{Alam:2012fw}) the angular separation for the U-shaped embeddings. The angular separation in this case is given by
\begin{eqnarray}
\cphi = 12 c \int_{r_0}^{r_*} \frac{dr}{e^{g}\sqrt{b}} \frac{1}{\sqrt{e^{2f + 6g+ 2\Phi} \left[ b  + h (B^2 b - e^2 ) \right]  - 6 c^2 }} \ ,
\end{eqnarray}
where 
\begin{eqnarray}
c = \left. \frac{1}{\sqrt{6}} e^{f+ 3g + \Phi} \left[ b + h \left( B^2 b - e^2 \right) \right]^{1/2} \right|_{r = r_0} \ .
\end{eqnarray}
It can be checked from the above expression that the large $r_0$ limit corresponds to large $c$ limit. Our task is to investigate what happens to this angular separation as $c \to \infty$. Using the solutions in (\ref{solb})-(\ref{solg}), it can be shown that in the limit $c \to \infty$ we get
\begin{eqnarray}
\cphi = \int_1^{r_*/r_0} dy \, \cI_1(y, \epsilon) -  \frac{1}{r_0^4} \left(e^2 - B^2\right) \int_1^{r_*/r_0} dy \, \cI_2 (y, \epsilon) + \ldots \ ,
\end{eqnarray}
where we have defined $y = r/ r_0$. Here we will not present the functional forms of $\cI_1$ or $\cI_2$ since they are not particularly illuminating; however, we will remark that the resulting integrals, performed over the variable $y$, are positive numbers. For large values of $r_0$, we can have two physically different regimes to consider: one where $r_* \gg r_0$ and one where $r_0 \sim r_*$. In the former regime, the angular separation tends to asymptote to a constant value and the behaviour is similar to what is discussed in \cite{Alam:2012fw}; however as $r_0$ increases, we enter the second regime and $\cphi \to 0$ eventually. Thus, in this case, irrespective of the relative magnitudes of the electric and the magnetic fields, there will always be a corresponding phase transition.

To make connection with our earlier work in \cite{Alam:2012fw} with a vanishing back-reaction, let us recall that there we found an upper limit on the electric field $e < B$, beyond which no phase transitions happen. We can recover this result from the above expression. To do so, let us set $\epsilon = 0$, which restores the conformal symmetry and we will have $r_*/r_0 \to \infty$ now. In that case, $\cphi < \sqrt{6} \pi/4$ for $e >B$ and therefore no phase transition takes place.

\subsection{Including a chemical potential}

There are two types of chemical potential we can introduce: U(1) (baryonic) and isospin\footnote{In this case, one will require two flavour branes. See {\it e.g.}~\cite{Parnachev:2007bc} introducing an isospin chemical potential in the Sakai-Sugimoto model.}. We can explore their effect in both the back-reacted and non-back-reacted backgrounds. We find that the effects are very nearly the same in any of these cases, {\it i.e.}~back-reaction does not significantly alter the results. Also an isospin chemical potential yields results that are qualitatively similar to the ones obtained in the U(1) case. Thus for simplicity, we will discuss the U(1) chemical potential case in the absence of back-reaction. 

The canonical way to realize a chemical potential is to excite the time-component of the gauge field $A_t(r)$, which will give rise to a bulk field strength $F_{tr}$. The corresponding DBI action takes the form
\begin{eqnarray}
&& S = \cN_T \int dr r^3 \left(1 + \frac{r^2}{6} b(r) \phi'^2 - a_t'^2 \right)^{1/2} = \cN_T \int dr \cL\ ,  \label{chemaction} \\
&& b(r) = 1 - \left(\frac{r_H}{r}\right)^4 \ , \quad a_t: = \left(2 \pi \alpha'\right) A_t  \ .
\end{eqnarray}
Once again we remind the reader that we are considering the case when the back-reaction vanishes, {\it i.e.}~$\epsilon = 0$ limit of the background in (\ref{solb})-(\ref{solg}).

The equations of motion that result from the above action are given by
\begin{eqnarray}
&& \left(\frac{r^2}{6}\right) \frac{r^3 b(r) \phi'}{\left( 1 + \frac{r^2}{6} b(r) \phi'^2 - a_t'^2  \right)^{1/2}} = c \ , \label{eomphi} \\
&& \frac{r^3 a_t'}{\left( 1 + \frac{r^2}{6} b(r) \phi'^2 - a_t'^2  \right)^{1/2}} = d \ , \label{eomat}
\end{eqnarray}
where we have seen the constant $c$ appear before and $d$ denotes a new constant of motion. Before going further, let us discuss these equations in more details. Note that, in order for the U-shaped profiles to join smoothly at some $r= r_0$, we need to impose $\phi'(r_0) \to \infty$.\footnote{Note that, if we relax the condition $\phi'(r_0) \to \infty$, then the physics is richer. One will have to consider including explicit sources which can support the radial field strength on the U-shaped profiles. Usually there are two candidates for such sources: a baryon vertex and a bunch of fundamental string attached to the probe brane at $r_0$. The qualitative picture is similar to \cite{Dymarsky:2010ci}, which analyzes baryons in the Klebanov-Strassler set-up\cite{Dymarsky:2009cm}. In the Klebanov-Witten case, we have one candidate for explicit source: a bunch of fundamental strings that stretch from $r=r_0$ to $r=r_H$. However, we will not discuss the physics when such explicit sources are included.} From the above equations of motion we can conclude that this condition leads to set $d=0$ identically, or demand that $a_t'(r_0) \to \infty$ as well. In order to keep the norm of the bulk vector field $F_{tr}$ finite, we conclude that for the U-shaped profiles we have $d=0$ and they will not be affected by the inclusion of the chemical potential. This is expected on physical grounds since the bulk radial field has nowhere to go for the U-shaped profiles. On the other hand, for the parallel embeddings there will be a non-trivial gauge field.

For the parallel shaped case, the asymptotic behaviour of the gauge field takes the form
\begin{eqnarray} \label{asympat}
a_t (r) \simeq (2\pi\alpha') \mu -  \frac{ 2d}{r^2} + \ldots \ , 
\end{eqnarray}
where the constant $\mu$ is related to the chemical potential and the constant $d$ is related to the charge density of the system. Let us denote the charge density in the probe sector by
\begin{eqnarray}
\rho = \frac{\delta S}{\delta F_{rt}}  \quad \implies \quad \rho = \left(4 \cN_T \right) \left( 2 \pi \alpha' \right) d \ , \label{chargeden}
\end{eqnarray}
where we have used the asymptotic expansion in (\ref{asympat}) and the definition of $S$ from (\ref{chemaction}). On the other hand, the chemical potential of the system can be obtained by
\begin{eqnarray}
\mu = \frac{1}{2\pi\alpha'} \int_{r_H}^{\infty} a_t'(r) dr = \frac{d}{(2\pi\alpha') r_H^2} \frac{\, _2F_1\left(\frac{1}{3},\frac{1}{2};\frac{4}{3};- (d^2/r_H^6) \right)}{2} \ ,
\end{eqnarray}
where we have utilized the solution of $a_t'(r)$ in terms of $d$ from equation (\ref{eomat}) and $_2F_1$ is a hypergeometric function. Our task henceforth will be to determine the favoured phase among the U-shaped and the parallel profiles. To determine this, we need to evaluate and compare the corresponding thermodynamic free energies of the individual phases. In this context, we can address this in two inequivalent ways, {\it i.e.}~in grand canonical and canonical ensembles respectively.

 \subsubsection{Grand Canonical Ensemble}

Let us first work in the grand canonical ensemble. If we take the action in (\ref{chemaction}) in its full generality and evaluate the corresponding variation on-shell, then we are left with the following boundary term
\begin{eqnarray}
\delta S = \cN_T \left[ \frac{\partial \cL}{\partial \phi'} \delta \phi + \frac{\partial \cL} {\partial a_t'} \delta a_t \right]_{r_{\rm min}}^{r_{\rm max}} \ ,
\end{eqnarray}
where $r_{\rm min}$ denotes the lowest point in the infrared, $r_{\rm max}$ denotes the UV-cutoff and $\cL$ denotes the Lagrangian density in (\ref{chemaction}). Using the equations of motion in (\ref{eomphi}) and (\ref{eomat}), we get
\begin{eqnarray}
&& \delta S = \cO_\phi \delta \left(\Delta\phi_{\infty} \right) + \cO_{\mu} \delta \mu \ , \\
&& \cO_\phi = \cN_T c \ , \quad \cO_{\mu} = \cN_T \left(2 \pi \alpha'\right) d \ ,
\end{eqnarray}
which implies that the natural thermodynamic variables in the corresponding ensemble are $\Delta\phi_{\infty}$ and $\mu$. This means that the on-shell action defines the corresponding Gibbs free energy in the grand canonical ensemble. The corresponding canonically conjugate variables are $c$, which is a condensate-like object, and $d$, which is related to the charge density {\it via} (\ref{chargeden}).

\begin{figure}[h!]
\centering
\includegraphics[scale=0.7]{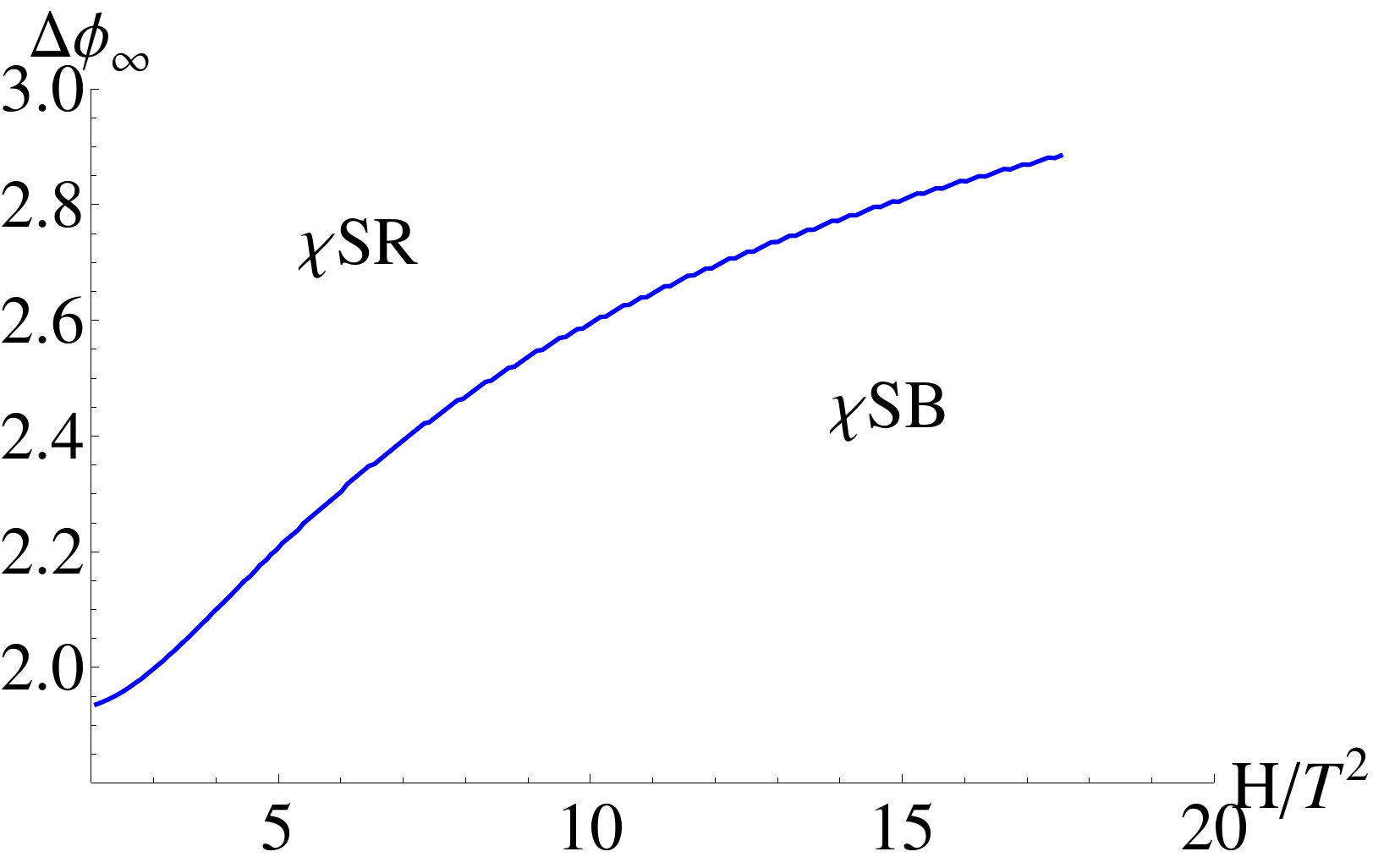}
\caption{The phase diagram in the non-backreacted background with a U(1) chemical potential for $(2\pi\alpha^{\prime})\mu = 1.0$ in the grand canonical ensemble.} 
\label{fig9}
\end{figure}

In the grand canonical ensemble it is straightforward to check that at finite temperature, the chiral symmetry restored phase is always favoured for any value of the chemical potential, which qualitatively is the same physics we observe at vanishing chemical potential as well \cite{Alam:2012fw}. In the absence of any back-reaction, we can only obtain a non-trivial phase structure after the inclusion of a magnetic field. This can be easily achieved as before, see {\it e.g.}~equation (\ref{magfield}). Introducing this field will deform the DBI Lagrangian and we will get
\begin{eqnarray}
\cL = r^3 \left(1 + \frac{r^2}{6} b(r) \phi'^2 - a_t'^2 \right)^{1/2}  \left(1 + B^2 L^4/r^4 \right)^{1/2} \ .
\end{eqnarray}
The analysis of the solutions proceeds as before and one can conclude that a non-trivial chemical potential will exist only for the parallel shaped case. For a fixed value of the chemical potential one can now explore the phase diagram, which is shown in fig.~\ref{fig9}. From the phase diagram it is clear that the effect of magnetic catalysis observed in \cite{Alam:2012fw} for vanishing chemical potential survives here. Here we have presented only a representative diagram, but have checked explicitly that the qualitative behaviour remains similar for a wide range of values for the chemical potential. 
Evidently, one can also explore the phase diagram (even without any magnetic field) resulting from the non-trivial dynamics once back-reaction is included. However we will not attempt this here, because, as should be noted, once the back-reaction is taken into account, the qualitative nature of fig.~\ref{fig9} will not change since we treat the back-reaction perturbatively. Thus, the physics will remain unchanged at the leading order.

\subsubsection{Canonical Ensemble}

\begin{figure}[h!]
\begin{center}
\subfigure[] {\includegraphics[angle=0,
width=0.48\textwidth]{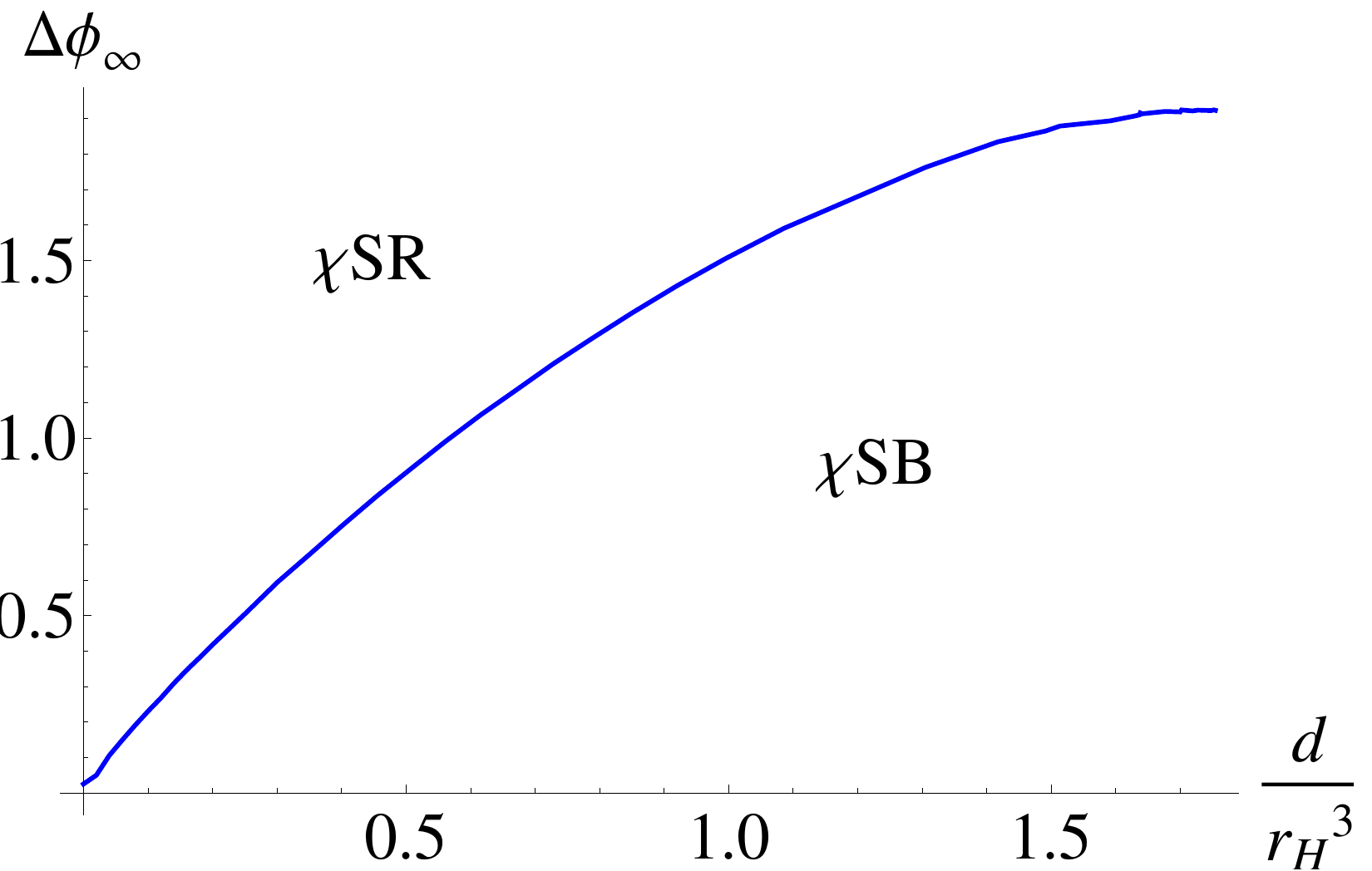} }
 \subfigure[] {\includegraphics[angle=0,
width=0.48\textwidth]{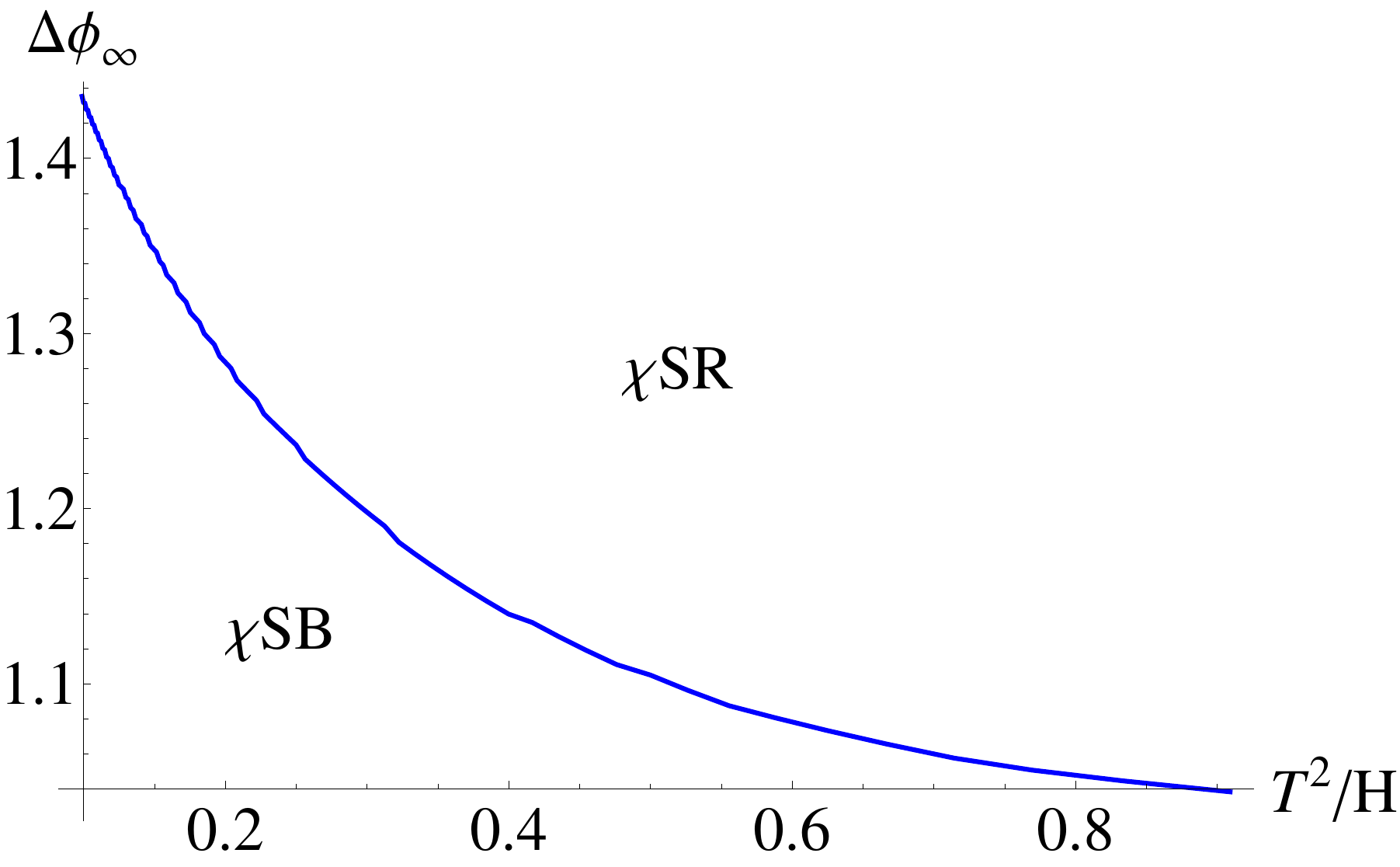} }\caption{Panel (a): Phase diagram at vanishing magnetic field, in the canonical ensemble. Note that there is no corresponding phase transition in the grand canonical ensemble. Panel (b): A representative phase diagram in the presence of a magnetic field. Here we have fixed $d=2$.}
\label{fig7}
\end{center}
\end{figure}

Let us now switch gear and discuss the physics in the canonical ensemble, which is characterized by the charge density rather than the chemical potential. The corresponding free energy is the so called Helmholtz free energy that can be obtained by a Legendre transformation of the Gibbs free energy. In terms of the on-shell action of the probe, we now need to consider the following quantity (see {\it e.g.}~\cite{Kobayashi:2006sb})
\begin{eqnarray}
\tilde{\cL} = \cL - \frac{\partial\cL}{\partial a_t'} a_t' \ .
\end{eqnarray}
Clearly, since the U-shaped profiles do not support a non-trivial $a_t'$, the Legendre transformation will change the corresponding free energy for the parallel shaped profiles only. It can be checked that in the canonical ensemble we have a non-trivial phase diagram when a non-zero charge density is introduced. We obtain a phase transition both at vanishing and at non-zero magnetic field. The corresponding diagram is shown in fig.~\ref{fig7}.
Moreover, we can also explore the physics in the presence of a magnetic field. A representative phase diagram in shown in fig.~\ref{fig8}, which again demonstrates the effect of magnetic catalysis that we have observed and commented on several times by now. Before concluding this section let us remark that the inclusion of the back-reaction again does not have a significant effect on the qualitative physics, as expected.

\section{Discussion of results}

In this article, we investigated various aspects of the back-reacted solution that we obtained in \cite{Ihl:2012bm}. In keeping with our perturbative approach, we find that the physics is qualitatively similar to the original background both w.r.t.~the back-reacted background and the additional probe sector. We observe the familiar magnetic catalysis effect\footnote{See {\it e.g.} \cite{Filev:2007gb, Albash:2007bk, Johnson:2008vna, Filev:2012ch} for similar effects in other models in the probe limit and in \cite{Filev:2011mt} beyond the probe limit.}, and the emergence of a pseudo-horizon in the probe sector. We also initiated a study of the corresponding phase structure introducing a chemical potential in this model.

There are various directions for future work. When the back-reaction is included, the presence of the phase transition in the additional probe sector demands a more thorough study of the model. This phase transition is perhaps signaling that we need to obtain backgrounds back-reacted by both the parallel and the U-shaped profiles and compare their energetics to decide whether the true picture is richer in physics. This is a rather intriguing possibility that we hope to address in the future.

For our current work, it was convenient to consider the additional probe sector and exciting various fields restricted to this sector. In principle, it should not be possible to distinguish between the additional probe sector and the back-reacting probe sector and thus what we have analyzed here is at best an approximate situation, however, it would be very instructive to consider the back-reaction including such worldvolume gauge fields: The back-reaction by a probe magnetic field will induce anisotropy in the system, a chemical potential will induce a charged black hole background, an electric field will induce a time-dependent background where the actual bulk event-horizon will be increasing with time. Such results would be exciting to further analyze and understand.

Note that our analysis of the model with the chemical potential is rather rudimentary in the sense that we did not include any source terms. We observed that the phenomenon of magnetic catalysis persists and is independent of the chemical potential. On the other hand, it has been observed in {\it e.g.}~\cite{Preis:2010cq} that an inverse magnetic catalysis effect exists for the Sakai-Sugimoto model. It will be very interesting to see whether it is possible to find a similar physical effect in this model once sources are included. We leave such interesting research opportunities for future work.

\section*{Acknowledgements}

We are grateful to Jacques Distler and Vadim Kaplunovsky for very useful comments. This material is based upon work supported by an IRCSET postdoctoral fellowship (MI), the National Science Foundation under Grant no.~PHY-0969020 (MSA, AK and SK) and by a Simons postdoctoral fellowship awarded by the Simons Foundation (AK). The work of SK is also supported by the Texas Cosmology Center, which is supported by the College of Natural Sciences and the Department of Astronomy at the University of Texas at Austin and the McDonald Observatory. AK would like to acknowledge the warm hospitality at the TIFR in Mumbai, IISc in Bangalore for providing a conducive environment during the final stages of this work. MI is grateful for the hospitality at Zewail City of Science and Technology, Giza, Egypt where the finishing touches were put on this work.



\end{document}